\documentclass[a4paper,11pt]{article}
\usepackage{graphicx,amssymb,amstext,amsmath,mathtools}
\usepackage{color}

\definecolor{cbl}{rgb}{0,0,1}                

\newcommand{\bc}{\begin{center}}
\newcommand{\ec}{\end{center}}
\def\ba#1{\begin{array}{#1}\displaystyle}
\newcommand{\ea}{\end{array}}

\newcommand{\beq}{\begin{equation}}
\newcommand{\eeq}{\end{equation}}
\newcommand{\beqa}{\begin{eqnarray}}
\newcommand{\eeqa}{\end{eqnarray}}

\newcommand{\bi}{\begin{itemize}}
\newcommand{\ei}{\end{itemize}}

\newcommand{\bra}{\langle}
\newcommand{\ket}{\rangle}

\newcommand{\Tr}{{\rm Tr}}

\newcommand{\TT}{{\cal T}}

\begin{document}
\begin{titlepage}
\vspace{0.2cm}
\begin{center}

{\large{\bf{Branch Point Twist Field Correlators in the Massive Free Boson Theory}}}

\vspace{0.8cm} {\large \text{Davide Bianchini}${}^{\bullet}$ and \text{Olalla A. Castro-Alvaredo}${}^{\circ}$}

\vspace{0.2cm}
{Department of Mathematics, City University London, Northampton Square EC1V 0HB, UK }
\end{center}

\vspace{1cm}
Well-known measures of entanglement in one-dimensional many body quantum systems, such as the entanglement entropy and the logarithmic negativity, may be expressed  in terms of the correlation functions of local fields known as branch point twist fields in a replica quantum field theory. In this ``replica'' approach the computation of measures of entanglement generally involves a mathematically non-trivial analytic continuation in the number of replicas. In this paper we consider two-point functions of twist fields and their analytic continuation in the 1+1 dimensional massive (non-compactified) free Boson theory. This is one of the few theories for which all matrix elements of twist fields are known so that we may hope to compute correlation functions very precisely. We study two particular two-point functions which are related to the logarithmic negativity of semi-infinite disjoint intervals and to the entanglement entropy of one interval. We show that our prescription for the analytic continuation  yields results which are in full agreement with conformal field theory predictions in the short-distance limit. We provide numerical estimates of universal quantities and their ratios, both in the massless (twist field structure constants) and the massive (expectation values of twist fields) theory. We find that particular ratios are given by divergent form factor expansions. We propose such divergences stem from the presence of logarithmic factors in addition to the expected power-law behaviour of two-point functions at short-distances. Surprisingly, at criticality these corrections give rise to a $\log(\log\ell)$ correction to the entanglement entropy of one interval of length $\ell$. This hitherto overlooked result is in agreement with results by Calabrese, Cardy and Tonni and has been independently derived by Blondeau-Fournier and Doyon (in preparation).
\medskip
\medskip

\vfill

\noindent 
${}^{\bullet}$ davide.bianchini@city.ac.uk\\
${}^{\circ}$ o.castro-alvaredo@city.ac.uk\\
 \hfill 
 
 \medskip
 \medskip
 \today

\end{titlepage}

\section{Introduction}
\begin{figure}[h!]
 \begin{center} 
 \includegraphics[width=14cm]{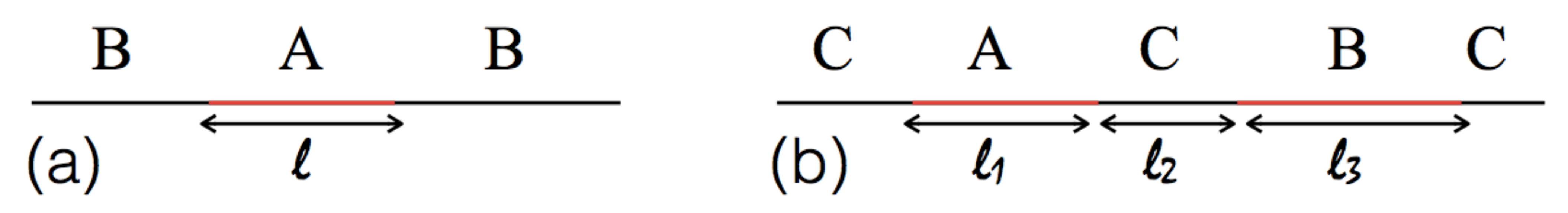} 
 \end{center} 
 \caption{Typical configurations for the entanglement entropy of one interval and the logarithmic negativity.} 
 \label{typical} 
 \end{figure}

The problem of quantifying the amount of entanglement which may be ``stored" in the ground state of a  many body quantum system has attracted the interest of the quantum information and theoretical physics communities for a long time. Measuring entanglement is of interest both if we are to employ entanglement as a quantum computing resource and if we want to learn more about the fundamental features of quantum states of highly complex quantum systems. Among such systems, 1+1-dimensional many body quantum systems have received considerable attention over the past decade. Much  work in this area has been inspired by the results  of Calabrese and Cardy \cite{Calabrese:2004eu} which used principles of Conformal Field Theory (CFT) to study a particular measure of entanglement, the entanglement entropy (EE) \cite{bennet}. In this seminal work, they generalised previous results \cite{HolzheyLW94} and provided theoretical support for numerical observations in critical quantum spin chains \cite{latorre1}. Before we proceed any further a few definitions are in order: let $|\Psi \rangle$ be a pure state describing the ground state of quantum spin chain at zero temperature. Consider a bi-partition of the chain such as in Fig.~\ref{typical}(a) (suppose there are periodic boundary conditions). Then the entanglement entropy associated to region $A$ may be expressed as $S(\ell)=-\Tr(\rho_A \log \rho_A)$ where $\rho_A=\Tr_B(|\Psi \rangle \langle\Psi|)$ is the reduced density matrix associated to subsystem $A$ and $\ell$ is the subsystem's length.

One of the main results of \cite{HolzheyLW94, latorre1, Calabrese:2004eu} describes the entanglement entropy of 1+1 dimensional many body quantum systems (such as spin chains) in the continuous limit at criticality. Such systems are described by CFT and their EE displays universal features expressed by the now famous formula: $S(\ell)=\frac{c}{3}\log \frac{\ell}{\epsilon}$. That is, the EE of a subsystem of length $\ell$ of an infinite critical system diverges logarithmically with the size of the subsystem, with a universal coefficient which is proportional to the central charge of the CFT, $c$. There are non-universal constant corrections to this leading behaviour  which may be encoded by a short-distance cut-off $\epsilon$. This behaviour has been numerically and analytically studied for a plethora of spin chain models in works such as \cite{latorre1,latorre2,latorre3,Jin,peschel,Lambert,Keating,Weston,fabian,parity}.
 
Another popular measure of entanglement is the logarithmic negativity (LN) \cite{VW,ple,erratumple,Eisert,Eisert2}. Consider again a quantum spin chain in a pure state $|\Psi \rangle$ and a partition such as depicted in Fig.~\ref{typical}(b). Then, the LN is a measure of the amount of entanglement between the two non-complementary sub-systems $A$ and $B$. Its formal definition depends on the reduced density matrix $\rho_{A\cup B}$ as $\mathcal{E}(\ell_1,\ell_2,\ell_3)=\log(\Tr|\rho_{A\cup B}^{T_B}|)$ where $T_B$ represents partial transposition with respect to subsystem $B$ and $|\rho|$ is the trace norm of $\rho$, that is the sum of the absolute values of its eigenvalues. The LN of 1+1 dimensional critical systems has been  studied numerically in \cite{negnum1,negnum2,negnum3} and more recently, both numerically and analytically exploiting fundamental conformal field theory principles, in  \cite{negativity1,negativity2}. Since then many particular models have been analysed (see e.g.~\cite{DeNobili,negativity3,negativity4,Alba}). However, for general configurations such as in Fig.~\ref{typical}(b) there is no known analytic formula for generic CFTs. There are however particular limits which are easier to treat such as the limit of adjoint intervals ($\ell_2 \rightarrow 0$) and the limit of semi-infinite disjoint intervals ($\ell_1, \ell_3 \rightarrow \infty$ keeping $\ell_2$ finite). The former has been studied in \cite{negativity1, negativity2} for generic CFT yielding the simple expression $\mathcal{E}(\ell_1,0,\ell_3)=\frac{c}{4}\log\frac{\ell_1\ell_3}{\epsilon(\ell_1+\ell_3)}$ whereas the latter is harder to treat in critical systems but is of interest in the study of quantum systems near criticality. Such systems are described by 1+1 dimensional massive quantum field theories which, unlike CFT, allow for the existence of a finite correlation length. The negativity of such systems was first studied in \cite{ourneg} where new results for both of the limits above in near-critical systems were obtained. 

In this paper we will be interested in a particular prescription for the calculation of both the EE of a single interval and the LN of semi-infinite disjoint regions. It turns out that both quantities may be expressed in terms of two-point functions of a particular class of fields known as branch point twist fields \cite{Calabrese:2004eu,entropy}. This relationship comes about through a technique commonly known as the ``replica trick''. The replica trick may be applied to both the computation of the EE and of the LN. It involves a rewriting of the definitions above as follows
\beq 
S(\ell)=-\lim_{n\rightarrow 1^+}\frac{d}{dn} \Tr(\rho_A^n) \qquad \text{and} \qquad \mathcal{E}(\ell_1,\ell_2,\ell_3)=\lim_{n_e\rightarrow 1^+} \log(\Tr(\rho_{A\cup B}^{T_B})^{n_e}), \label{rtrick}
\eeq 
where the symbol $n_e$ in the second formula means $n$ even, that is the limit $n$ to 1 must be carried out by analytically continuing the function from even, positive values of $n$ to $n=1$. The representations above were used first in \cite{HolzheyLW94} for the EE and in \cite{negativity1, negativity2} for the LN. The advantage of such representations is that both $\Tr(\rho_A^n)$ and $\Tr(\rho_{A\cup B}^{T_B})^{n_e}$ admit a natural physical interpretation as partition functions in ``replica'' theories. The replica theory is a new model consisting of $n$ non-interacting copies of the original theory. In this context it is natural for $n$ to take positive integer values. However, the definitions (\ref{rtrick}) require that such traces be analytically continued from $n$ integer (and in the LN case, also even) to $n$ real and positive. Hence, formulae (\ref{rtrick}) are advantageous in that partition functions in replica theories may be computed systematically by various approaches, but also disadvantageous because the analytic continuations involved are often very difficult to perform and there is no generic proof of existence and uniqueness. 

It was first noted in \cite{Calabrese:2004eu} that the function $\Tr(\rho_A^n)$ may be expressed as a two-point function of fields with conformal dimension given by
\beq 
\Delta_n=\frac{c}{24}\left(n-\frac{1}{n}\right). 
\eeq 
In fact such fields had been previously discussed in the context of the study of orbifold CFT where they emerge naturally as symmetry fields associated to the permutation symmetry of the theory \cite{kniz,orbifold}. In \cite{entropy} such fields were named branch point twist fields and studied in the context of 1+1 dimensional massive QFT. Their connection to the cyclic permutation symmetry of the replica theory was made explicit by formulating their exchange relations with the fundamental fields of a generic replica QFT. For integrable QFT this allowed for the formulation of twist field form factor equations whose solutions are matrix elements of twist fields. Let $\mathcal{T}$ be a twist field associated to the cyclic permutation symmetry $j \mapsto j+1$ and $\tilde{\mathcal{T}}$ its conjugate, associated with the permutation symmetry $j \mapsto j-1$ with $j=1, \ldots, n$. Then, we may write:
\beq 
\Tr(\rho_A^n)=\epsilon^{4\Delta_n}\langle \mathcal{T}(0)\tilde{\mathcal{T}}(\ell) \rangle_n \qquad \text{and} \qquad \Tr(\rho_{A\cup B}^{T_B})^{n}=\epsilon^{8\Delta_n}\langle \mathcal{T}(-\ell_1)\tilde{\mathcal{T}}(0)\tilde{\mathcal{T}}(\ell_2) {\mathcal{T}}(\ell_2+\ell_3)\rangle_n. \label{TTTT}
\eeq 
At criticality, these formulae may be used directly to derive the expressions for $S(\ell)$ and $\mathcal{E}(\ell_1,0,\ell_2)$ given above. The same formulae may be used to study QFT beyond criticality as done in \cite{entropy,ourneg}. In this paper we will analyse the short-distance (e.g. $\ell \ll 1$) behaviour of the correlators $\langle \mathcal{T}(0)\tilde{\mathcal{T}}(\ell) \rangle_n$ and $\langle \mathcal{T}(0)\mathcal{T}(\ell) \rangle_n=\langle \tilde{\mathcal{T}}(0)\tilde{\mathcal{T}}(\ell) \rangle_n$ in a massive free Boson theory. At short-distances we expect the massive QFT to be described by its corresponding ultraviolet limit (that is, the massless (non-compactified) free Boson CFT). Thus, we expect these two-point functions to exhibit power-law behaviours with powers related to the dimension of twist fields. 
Extracting these short-distance behaviours from a form factor expansion (which is eminently a large-distance expansion) is generally highly non-trivial and can seldom be done with precision for any fields. However, as we will see, this can be done with great precision for the massive free Boson, on account of the theory's simplicity and the special properties of the twist field form factors. For the massive free Boson all form factors of twist fields, that is objects such as
\beq 
F_{k}^{\mathcal{T}|j_1\ldots j_k}(\theta_1, \cdots, \theta_k;n):=\langle 0|\mathcal{T}(0)|\theta_1,\cdots, \theta_k \rangle_{j_1\ldots j_k}/{\bra \TT\ket_n}, \label{ff}
\eeq 
are known explicitly. Here $\langle 0|$ represents the vacuum state and $|\theta_1,\cdots, \theta_k \rangle_{j_1\ldots j_k}$ represents an in-state of $k$ particles with rapidities $\theta_1, \dots, \theta_k$ and quantum numbers $j_1\ldots j_k$. In the free Boson case, these quantum numbers are just the copy number of the Boson in the replica theory. Here we have chosen to normalise all form factors by a constant (the vacuum expectation value of the twist field $\bra \TT\ket_n$). This will be convenient for later computations. 

By reconstructing the short-distance (power-law) behaviour of the correlators $\langle \mathcal{T}(0)\tilde{\mathcal{T}}(\ell) \rangle_n$ and $\langle \mathcal{T}(0)\mathcal{T}(\ell) \rangle_n$ for $n\geq 1$, integer or not, we will provide strong evidence for our approach to performing the analytic continuation of the correlators in $n$. This will provide support for our methodology and will allow us to examine twist field two-point functions further and extract values of universal quantities such as expectation values and structure constants of twist fields.

\medskip 

The paper is organized as follows: In sections 2 and 3 we review basic CFT and QFT results, regarding the short distance behaviour of two-point functions of twist fields and how these two-point functions may be expressed in terms of the form factors (\ref{ff}). In section 4 we show how the power-law decay of two-point functions of twist fields may be obtained exactly from form factors in the massive free Boson theory for $n\geq 1$ real. In section 5 we provide form factor expansions for the constant (universal) coefficients that multiply the leading power-law in the two-point functions of twist fields. We employ these expansions to obtain numerical predictions for the ratio of the structure constant $\mathcal{C}_{\TT\TT}^{\TT^2}$ and the expectation value $\bra \TT \ket_n$, analytically continued from $n$ odd and for the structure constant $\mathcal{C}_{\TT\TT}^{\TT^2}$ analytically continued from $n$ even. We compare our values of $\mathcal{C}_{\TT\TT}^{\TT^2}$ for $n$ even to analytical values obtained in \cite{negativity2} and find good agreement. We numerically examine the limit $\lim_{n_e \rightarrow 1^{+}} \mathcal{C}_{\TT\TT}^{\TT^2}$ and compare to an analytical prediction given in \cite{negativity2}. In section 6 we present an interpretation of the emergence of divergent sums in the representation of particular ratios of expectation values and structure constants of the massive free Boson theory. We propose that such divergences must be related to the presence of logarithmic corrections to the two-point functions at criticality. We conclude that such corrections will give rise to an additional $\log(\log\ell)$ term in the EE and the R\'enyi entropy of one interval in the massless (non-compactified) free Boson theory. This is in full agreement with previous results for the LN \cite{negativity2} and the EE \cite{log} of the compactified massless free Boson in the limit of infinite compactification ratio. For the EE the presence of such corrections has also been established analytically by a different method in \cite{Part2} but had been overlooked in \cite{callan}.
In section 7 we compare  our numerical estimates of the value of $\lim_{n_e \rightarrow 1^{+}} \mathcal{C}_{\TT\TT}^{\TT^2}$  as well as the analytical value given in \cite{negativity2} to a value that can be read off from numerical results in \cite{EZ} for the LN of a harmonic chain out of equilibrium and their CFT interpretation \cite{MB}. We present our conclusions in section 8. Appendix A collects some useful summation formulae which feature in the form factor expansions of sections 4 and 5. Appendix B provides a discussion and assessment of the error of  some of our numerical procedures.  

\section{Conformal Field Theory Recap}
As described in the introduction, we wish to study the two-point functions  $\langle \mathcal{T}(0)\tilde{\mathcal{T}}(\ell) \rangle_n$ and $\langle \mathcal{T}(0)\mathcal{T}(\ell) \rangle_n$ and examine their short-distance behaviour. This behaviour is entirely predicted by CFT and may be expressed as 
\beq
\log\left(\frac{\langle \TT(0)\tilde{\TT}(\ell) \rangle_n}{\langle \TT \rangle_n^2}\right)_{m\ell \ll 1}=-4\Delta_n \log \ell -2 \log \bra \TT \ket_n. \label{log1}
\eeq 
Similarly 
\beq 
\log\left(\frac{\langle \TT(0){\TT}(\ell) \rangle_n}{\langle \TT \rangle_n^2}\right)_{m\ell \ll 1}=\left\lbrace\begin{array}{cc}
-2\Delta_n \log\ell+\log\frac{\mathcal{C}_{\TT \TT}^{\TT^2}}{\langle \TT \rangle_n}& \text{for n odd}\\
-4(\Delta_n-\Delta_{\frac{n}{2}})\log\ell+\log\frac{\langle \TT \rangle_{\frac{n}{2}}^2\mathcal{C}_{\TT \TT}^{\TT^2}}{\langle \TT \rangle_n^2}& \text{for n even}\\
\end{array}\right.\label{log2}
\eeq 
Note that by examining the next-to-leading order ($\ell$-independent) corrections above we may extract values for universal QFT quantities such as the twist field expectation value $\bra \TT\ket_n$ and the structure constants $\mathcal{C}_{\TT\TT}^{\TT^2}$ and their ratios. These are difficult to compute by other methods, demonstrating once more that the form factor approach in particularly powerful in this context. 

The difference between the $n$ odd and $n$ even cases was first discussed in \cite{negativity1, negativity2} and follows from the leading term in the conformal OPE of the field $\TT$ with itself, which takes the form
\beq 
\TT(0){\TT}(\ell) \sim C_{\TT\TT}^{\TT^2} \ell^{-4\Delta_n+2\Delta_{\TT^2}} \TT^2(0)+\cdots  
\eeq 
This leading term is characterized by a new twist field $\TT^2$ of conformal dimension $\Delta_{\TT^2}$ which is associated with the permutation symmetry $j \mapsto j+2$ for $j=1,\ldots, n$. As discussed in \cite{negativity1, negativity2} the nature of this field is very different depending on whether $n$ is odd or even. Whereas for $n$ odd, the field $\TT^2$ is equivalent to the field $\TT$ (the permutation $j\mapsto j+2$ still allows for visiting all copies, albeit in a different order), for $n$ even the permutation $j \mapsto j+2$ divides even- and odd-labeled copies so that $\TT^2$ is equivalent to two copies of $\TT$ acting on a $\frac{n}{2}$-replica theory. Consequently the conformal dimension of $\TT^2$ is $\Delta_{\TT^2}=\Delta_n$ for $n$ odd and $\Delta_{\TT^2}=2\Delta_{\frac{n}{2}}$ for $n$ even. For the same reasons $\bra\TT^2\ket_n=\bra \TT\ket_n$ for $n$ odd and $\bra\TT^2\ket_n=\bra \TT\ket_{\frac{n}{2}}^2$ for $n$ even. This simple interpretation also shows how the analytic continuations (\ref{rtrick}) from $n$ even and $n$ odd should be different. Note that, $\bra \TT \ket_1=1$ both for massive and massless theories as the twist field becomes the identity field at $n=1$.

In massive theories, the correlator $\bra \TT(0) \TT(\ell)\ket_n$ encodes the $\ell$-dependent part of the negativity $\mathcal{E}(\infty, \ell, \infty)$ of semi-infinite disjoint regions. This follows simply from the definition (\ref{TTTT}) and the factorization of correlation functions at large distances in massive QFT.

In this paper we will use a form factor expansion of these correlators to extract the leading term (the $\log \ell$ term). We will turn our attending to the next-to-leading order corrections in section~5. 

\section{Form Factor Expansion of two-Point Functions}
In a massive integrable QFT such as the massive free Boson, the functions (\ref{log1})-(\ref{log2}) admit a natural large $m\ell$ expansion in terms of form factors. In general we have that the (normalized) logarithm of the two-point function of local fields $\mathcal{O}_1, \mathcal{O}_2$ admits and expansion of the form 
\beq 
\log\left(\frac{\bra \mathcal{O}_1(0)\mathcal{O}_2(\ell) \ket}{\bra \mathcal{O}_1 \ket \bra \mathcal{O}_2 \ket} \right)= \sum_{j=1}^\infty c^{\mathcal{O}_1 \mathcal{O}_2}_j(\ell), \label{cs}
\eeq 
with
\beqa 
c^{\mathcal{O}_1\mathcal{O}_2}_j(\ell)&=&\frac{1}{j!(2\pi)^{j} } \sum_{p_1,\ldots, p_{j}=1}^N \int_{-\infty}^\infty d\theta_1 \cdots \int_{-\infty}^\infty d\theta_{j} \, \, {h}^{\mathcal{O}_1\mathcal{O}_2|p_1\ldots p_{j}}_{j}(\theta_1,\cdots,\theta_{j}) e^{-m\ell \sum_{i=1}^{j}\cosh\theta_i},
\label{genc}
\eeqa 
where the functions $h_j^{\mathcal{O}_1\mathcal{O}_2|p_1\ldots p_j}(\theta_1,\cdots, \theta_j)$ are given in terms of the form factors of the fields involved, $N$ is the number of particles in the spectrum and $p_i$ represent the particle's quantum numbers. For example:
\beqa 
h_1^{\mathcal{O}_1\mathcal{O}_2|p}(\theta)&=& F_1^{\mathcal{O}_1|p_1}(\theta) (F_1^{\mathcal{O}_2^\dagger|p_1}(\theta))^*\nonumber\\
h_2^{\mathcal{O}_1\mathcal{O}_2|p_1 p_2}(\theta_{1},\theta_2)&=&{F}^{\mathcal{O}_1|p_1 p_2}_2(\theta_{1},\theta_{2})({F}^{\mathcal{O}_2^\dagger|p_1 p_2}_2(\theta_1,\theta_2))^*-h_1^{\mathcal{O}_1\mathcal{O}_2|p_1}(\theta_1)h_1^{\mathcal{O}_1\mathcal{O}_2|p_2}(\theta_2), \label{exh}
\eeqa 
and so on. Here we have used the generic property:
\beq 
\bra \theta_j \ldots \theta_1|\mathcal{O}_2(0)|0\ket=\bra 0 |\mathcal{O}_2^\dagger(0)|\theta_1\ldots\theta_j\ket^* =:F_j^{\mathcal{O}_2^\dagger|p_1\ldots p_j}(\theta_1,\ldots,\theta_j)^*. \label{conjugation}
\eeq 
The expansion (\ref{genc}) with (\ref{exh}) is usually referred to as the cumulant expansion of the two-point function (see e.g.~\cite{Smir,takacs,karo}) and it is particularly well suited to extract the leading $\log \ell$ behaviours seen earlier. If all form factors are know, this may be done by employing the fact that the operators $\mathcal{O}_1, \mathcal{O}_2$ are spinless (this will be the case for twist fields) and thus relativistic invariance implies that all form factors depend only on rapidity differences. In other words, one of the rapidities in the integrals (\ref{genc}) may be integrated over leading to 
\beqa 
c^{\mathcal{O}_1\mathcal{O}_2}_j(\ell)&=&\frac{2}{j!(2\pi)^{j} } \sum_{p_1,\ldots, p_{j}=1}^N \int_{-\infty}^\infty d\theta_2 \cdots \int_{-\infty}^\infty d\theta_{j} \, \, {h}^{\mathcal{O}_1\mathcal{O}_2|p_1\ldots p_{j}}_{j}(0,\theta_2,\cdots,\theta_{j}) K_0(m\ell d_j),
\label{genc222}
\eeqa 
where $K_0(x)$ is a Bessel function and 
\beq 
d_j^2=\left(\sum_{p=2}^j \cosh \theta_p +1\right)^2-\left(\sum_{p=2}^j \sinh \theta_p\right)^2. \label{dj}
\eeq 
Provided the functions ${h}^{p_1\ldots p_{j}}_{j}(0,\theta_2,\cdots,\theta_{j})$ vanish for large $\theta$ we may, for $m\ell \ll 1$ expand the Bessel function as $K_0(m\ell d_j)=-\log\ell-\gamma+\log 2-\log (m d_j)+\cdots$ where $\gamma=0.5772157...$ is the Euler-Mascheroni constant. 
For $m\ell \ll 1$ we expect the behaviour
\beq 
\log\left(\frac{\bra \mathcal{O}_1(0)\mathcal{O}_2(\ell) \ket}{\bra \mathcal{O}_1 \ket \bra \mathcal{O}_2 \ket} \right)_{m\ell \ll 1}= -x_{\mathcal{O}_1\mathcal{O}_2}\log\ell-
K_{\mathcal{O}_1\mathcal{O}_2},\label{o1o2}
\eeq 
then, considering the leading term in the Bessel function expansion and summing the resulting series from (\ref{cs}) we have that
\beqa
x_{\mathcal{O}_1\mathcal{O}_2}=\sum_{j=1}^\infty \frac{2}{j!(2\pi)^{j}} \sum_{p_1,\ldots, p_{j}=1}^N \int_{-\infty}^\infty d\theta_2 \cdots \int_{-\infty}^\infty d\theta_{j} \, {h}^{\mathcal{O}_1\mathcal{O}_2|p_1\ldots p_{j}}_{j}(0,\theta_2,\cdots,\theta_{j}).\label{del}
\eeqa
In addition, the next-to-leading correction for small $m\ell$ can also be obtained as shown in \cite{karo} and is given by
\beqa
K_{\mathcal{O}_1 \mathcal{O}_2}&=&
\sum_{j=1}^\infty\frac{2}{j! (2\pi)^{j}} \sum_{p_1,\ldots,p_j=1}^N 
\int_{-\infty}^\infty d\theta_2 \cdots \int_{-\infty}^\infty d\theta_{j} \, h_j^{\mathcal{O}_1\mathcal{O}_2|p_1,\ldots, p_j}(0,\theta_2, \cdots, \theta_j)(\log\frac{m d_j}{2}+\gamma)\nonumber\\ 
&=&x_{\mathcal{O}_1\mathcal{O}_2}(\log\frac{m}{2}+\gamma)\nonumber\\
&& +\sum_{j=1}^\infty\frac{2}{j! (2\pi)^{j}} \sum_{p_1,\ldots,p_j=1}^N 
\int_{-\infty}^\infty d\theta_2 \cdots \int_{-\infty}^\infty d\theta_{j} \, h_j^{\mathcal{O}_1\mathcal{O}_2|p_1,\ldots, p_j}(0,\theta_2, \cdots, \theta_j)\log{d_j}\label{vev}
\eeqa
\subsection{Form Factors in the Massive Free Boson Theory}
It is now easy to adapt the definitions above to the two-point functions of interest. In our case we are considering a free Boson theory in a replica theory, so the particle number is $N=n$, where $n$ is the number of replicas. The form factors of free Boson twist fields were first reported in \cite{ourneg} and they can be expressed in terms of the two-particle form factor
\beq 
F_2^{\TT|11}(\theta_1,\theta_2;n)=\frac{\sin\frac{\pi}{n}}{2n\sinh\left(\frac{i\pi-\theta_1+\theta_2}{2n}\right)\sinh\left(\frac{i\pi+\theta_1-\theta_2}{2n}\right)}=F_2^{\tilde{\TT}|11}(\theta_1,\theta_2;n)
\eeq 
For simplicity we will from now on call
\beq 
f(\theta_1-\theta_2;n):=F_2^{\TT|11}(\theta_1,\theta_2;n). \label{2pff}
\eeq 
Form factors associated to other copy numbers can be simply obtained by employing the properties:
\beq 
F_2^{\TT|p_1 p_2}(\theta;n)=f(-\theta+ 2\pi i (p_2-p_1);n), \qquad
{F}_2^{\tilde{\TT}|p_1 p_2}(\theta;n)={f}(\theta+ 2\pi i (p_2-p_1);n). \label{r2}
\eeq 
A direct consequence of these properties is that for the free Boson ${F}_2^{\tilde{\TT}|p_1 p_2}(\theta;n)^*=F_2^{\TT|p_1 p_2}(\theta;n)$ since $F_2^{\TT|p_1 p_2}(\theta;n)=F_2^{\TT|p_1 p_2}(-\theta;n)$ as the scattering matrix is 1. A detailed derivation of (\ref{r2}) may be found in \cite{entropy,review}. Similar properties can also be derived for higher particle form factors, so that every form factor of $\tilde{\TT}$ may be ultimately expressed in terms of form factors of $\TT$ involving only particles in copy 1 of the theory \cite{review}.
In addition, due to the $\mathbb{Z}_2$ symmetry of the free Boson Lagrangian, there are only non-vanishing even-particle form factors. 
Higher even-particle form factors may be simply obtained by employing Wick's theorem. In general they are given by \cite{ourneg}
\beq 
F_{2j}^{\TT|11\ldots 1}(\theta_1,\ldots,\theta_{2j};n)=\sum_{\sigma\in S_{2j}} f(\theta_{\sigma(1) \sigma(2)};n)\cdots f(\theta_{\sigma({2j-1})\sigma(2j)};n), \label{higher}
\eeq
where $S_{2j}$ represents the set of all permutations of $\{1,\ldots,2j\}$ and $\theta_{ij}:=\theta_i-\theta_j$ (a function with this combinatorial structure is know as a permanent in mathematics). For example:
\beq 
F_{4}^{\TT|1111}(\theta_1,\theta_2,\theta_3,\theta_{4};n)=
f(\theta_{12};n)f(\theta_{34};n)+f(\theta_{13};n)f(\theta_{24};n)+f(\theta_{14};n)f(\theta_{23};n).
\eeq 
This formula can be easily generalised to generic particles (e.g. particles living in different replicas) by using the relations (\ref{r2}). 
\subsection{Form Factor Expansions in the Massive Free Boson Theory}
Following the definitions above, let us write
\beq 
\log\left(\frac{\bra \TT(0)\tilde{\TT}(\ell) \ket_n}{\bra \TT\ket_n^2} \right)= \sum_{j=1}^\infty c^{\TT\tilde{\TT}}_{2j}(\ell,n) \quad \text{and} \quad \log\left(\frac{\bra \TT(0){\TT}(\ell) \ket_n}{\bra \TT\ket_n^2} \right)= \sum_{j=1}^\infty c^{\TT{\TT}}_{2j}(\ell,n), \label{cs2}
\eeq 
with
\beqa 
c^{\TT\tilde{\TT}}_{2j}(\ell,n)&=&\frac{1}{(2j)!(2\pi)^{2j} } \sum_{p_1,\ldots, p_{2j}=1}^n \int_{-\infty}^\infty d\theta_1 \cdots \int_{-\infty}^\infty d\theta_{2j} \, \, {h}^{\TT\tilde{\TT}|p_1\ldots p_{2j}}_{2j}(\theta_1,\cdots,\theta_{2j}) e^{-m\ell \sum\limits_{i=1}^{2j}\cosh\theta_i},
\label{genc22}
\eeqa 
and
\beqa 
c^{\TT{\TT}}_{2j}(\ell,n)&=&\frac{1}{(2j)!(2\pi)^{2j} } \sum_{p_1,\ldots, p_{2j}=1}^n \int_{-\infty}^\infty d\theta_1 \cdots \int_{-\infty}^\infty d\theta_{2j} \, \, {h}^{\TT\TT|p_1\ldots p_{2j}}_{2j}(\theta_1,\cdots,\theta_{2j}) e^{-m\ell \sum\limits_{i=1}^{2j}\cosh\theta_i},
\label{genc3}
\eeqa
We now have all the formulae necessary to write down the functions $c^{\TT\TT}_{2j}(\ell;n)$ and $c_{2j}^{\TT\tilde\TT}(\ell;n)$ corresponding to (\ref{genc}) for the correlators (\ref{log1})-(\ref{log2}). The simple structure of the form factors (\ref{higher}) coupled with the nature of the cumulant expansion (\ref{cs}) leads to great simplifications of the functions $h_{2j}^{\TT\TT|p_1\ldots p_{2j}}
(\theta_1,\ldots,\theta_{2j})$ and $h_{2j}^{\TT\tilde{\TT}|p_1\ldots p_{2j}}
(\theta_1,\ldots,\theta_{2j})$ which are unique for free theories and have already been observed in previous work for the massive free Fermion \cite{nexttonext}.

As in the examples (\ref{exh}), the first term contributing to each function $h_{2j}^{\TT\tilde{\TT}|p_1\ldots p_{2j}}
(\theta_1,\ldots,\theta_{2j})$ takes the form 
\beqa 
&& F_{2j}^{\TT|p_1 \ldots p_{2j}}(\theta_1\cdots \theta_{2j};n)({F}_{2j}^{\TT|p_1\ldots p_{2j}}(\theta_1\cdots \theta_{2j};n))^*\nonumber\\
&& =\left(\sum_{\sigma\in S_{2j}} F^{\TT|p_1 p_2}_2(\theta_{\sigma(1)},
\theta_{\sigma(2)};n)\cdots F_2^{\TT|p_{2j-1}p_{2j}}(\theta_{\sigma({2j-1})}, \theta_{\sigma(2j)};n)\right)\nonumber\\
&& \times \left(\sum_{\sigma\in S_{2j}} (F^{\TT|p_1 p_2}_2(\theta_{\sigma(1)}, \theta_{\sigma(2)};n))^*\cdots (F_2^{\TT|p_{2j-1}p_{2j}}(\theta_{\sigma({2j-1})},\theta_{\sigma(2j)};n))^*\right),\label{suma}
\eeqa 
where we uses the fact that $\TT=\tilde{\TT}^\dagger$, that is equation (\ref{conjugation}) and the definition (\ref{higher}). Employing the same equations,
the first term contributing to the function $h_{2j}^{\TT{\TT}|p_1\ldots p_{2j}}
(\theta_1,\ldots,\theta_{2j})$ takes the form
\beqa  
&&{F}_{2j}^{\TT|p_1 \ldots p_{2j}}(\theta_{1}\ldots \theta_{2j};n)({F}^{\tilde{\TT}|p_1 \ldots p_{2j}}_{2j}(\theta_{1} \ldots \theta_{2j};n)^*=\left({F}_{2j}^{\TT|p_1 \ldots p_{2j}}(\theta_{1}\ldots \theta_{2j};n)\right)^2\nonumber\\ 
&&=\left(\sum_{\sigma\in S_{2j}} F_2^{\TT|p_1p_2}(\theta_{\sigma(1)},
\theta_{\sigma(2)};n)\cdots F_2^{\TT|p_{2j-1}p_{2j}}(\theta_{\sigma({2j-1})},\theta_{\sigma(2j)};n)\right)^2.\label{sumb}
\eeqa
where the second equality follows from generalising equations (\ref{r2}) to higher particle form factors to show that ${F}^{\tilde{\TT}|p_1 \ldots p_{2j}}_{2j}(\theta_{1} \ldots \theta_{2j};n)=({F}^{{\TT}|p_1 \ldots p_{2j}}_{2j}(\theta_{1} \ldots \theta_{2j};n)^*$. 
Each sum $\sum_{\sigma\in S_{2j}}$ above consists of $\frac{(2j)!}{2^j \, j!}$ terms. Therefore, their product will generate a sum of $(\frac{(2j)!}{2^j \, j!})^2$ terms. However, many of these terms are identical up to integration in all rapidities. For example, the sum (\ref{suma}) for $j=2$ is 
\beqa 
&& F_{4}^{\TT|p_1 p_2 p_3 p_4}(\theta_1,\theta_2,\theta_3,\theta_{4};n)({F}_{4}^{\TT|p_1 p_2 p_3 p_{4}}(\theta_1, \theta_2,\theta_3,\theta_{4};n))^*\nonumber\\
&& = \left[F_2^{\TT|p_1 p_2}(\theta_{1},\theta_2;n)F_2^{\TT|p_3 p_4}(\theta_3,\theta_4;n)+F_2^{\TT|p_1 p_3}(\theta_1,\theta_3;n)F_2^{\TT|p_3 p_4}(\theta_3, \theta_4;n)\right.\nonumber\\
&&\left. \qquad +F_2^{\TT|p_1 p_4}(\theta_1,\theta_4;n)F_2^{\TT|p_2 p_3}(\theta_2,\theta_3;n)\right]\left[F_2^{\TT|p_1 p_2}(\theta_{1},\theta_2;n) F_2^{\TT|p_3 p_4}(\theta_3,\theta_4;n)\right.\nonumber\\
&& \qquad \left.+F_2^{\TT|p_1 p_3}(\theta_1,\theta_3;n)F_2^{\TT|p_3 p_4}(\theta_3, \theta_4;n) +F_2^{\TT|p_1 p_4}(\theta_1,\theta_4;n) F_2^{\TT|p_2 p_3}(\theta_2,\theta_3;n)\right]^*\nonumber
\\
&& =_{\text{int}} 6 F_2^{\TT|p_1 p_2}(\theta_{1},\theta_2;n)(F_2^{\TT|p_2 p_3}(\theta_2,\theta_3;n))^* F_2^{\TT|p_3 p_4}(\theta_3,\theta_4;n)(F_2^{\TT|p_1 p_4}(\theta_1,\theta_4;n))^*  \nonumber\\
&& \qquad +3\left|F_2^{\TT|p_1 p_2}(\theta_{1},\theta_2;n)\right|^2\left|F_2^{\TT|p_1 p_2}(\theta_{3},\theta_4;n)\right|^2,\label{3terms}
\eeqa 
where the symbol $=_{\text{int}}$ means equality under integration in all rapidities. Employing the properties (\ref{r2}) this may be written as
\beqa 
&& F_{4}^{\TT|p_1 p_2 p_3 p_4}(\theta_1,\theta_2,\theta_3,\theta_{4};n)({F}_{4}^{\TT|p_1 p_2 p_3 p_{4}}(\theta_1, \theta_2,\theta_3,\theta_{4};n))^*\nonumber\\
&& =_{\text{int}} 6 f(\theta_{12}^{p_1-p_2};n)f((-\theta_{23})^{p_2-p_3};n)^* f(\theta_{34}^{p_3-p_4};n)f((-\theta_{14})^{p_1-p_4};n))^*\nonumber\\
&& \qquad +3\left|f(\theta_{12}^{p_1-p_2};n)\right|^2\left|f(\theta_{34}^{p_3-p_4};n)\right|^2, \label{6terms}
\eeqa 
where $\theta^{p}:=\theta+2 \pi i p$. Finally, for the free Boson we also have that $f(x^{p};n)^*=f((-x)^p;n)$. In general, it is easy to show that there are exactly $(2j-1)!$ terms (identical under integration) which are so-called ``fully-connected''. In the $j=2$ example above there are exactly 6 such terms, those in the first line of (\ref{3terms}). Including the sum over all indices $p_i$, these are terms of the form
\beq 
n\sum_{p_1,\ldots,p_{2j-1}=0}^{n-1} \left(f((-\theta_{12})^{p_1};n)\prod_{k=1}^{j-1} f(\theta_{2k+1\, 2k+2}^{p_{2k}-p_{2k+1}};n) \right)\left(f(\theta_{1\,2j}^{p_{2j-1}};n) \prod_{k=1}^{j-1} f(\theta_{2k\,2k+1}^{p_{2k}-p_{2k-1}};n) \right), \label{summ}
\eeq 
In (\ref{summ}) one sum has been carried out by simply setting $p_{2j}=1$ multiplying by a factor $n$ (since all copies are identical) and shifting all $p_i's$ by 1. The crucial observation is that all terms which are not of this form (that is, they factorise into separate multiple sums such as the terms in the last line of (\ref{3terms})) are cancelled in the cumulant expansions (\ref{exh}). They generate precisely the products of $h$-functions on the r.h.s. of each definition. In summary, combining (\ref{suma})-(\ref{sumb}) with the properties (\ref{r2}) and employing the symmetry properties induced by the integrals in (\ref{genc22}) and (\ref{genc3}), we find that
\beqa 
c^{\TT\tilde{\TT}}_{2j}(\ell,n)&=&\frac{n}{(2j)(2\pi)^{2j} } \sum_{p_1,\ldots, p_{2j-1}=0}^{n-1} \int_{-\infty}^\infty d\theta_1 \cdots \int_{-\infty}^\infty d\theta_{2j} \,e^{-m\ell \sum_{i=1}^{2j}\cosh\theta_i} \nonumber\\ 
&& \times \left(f((-\theta_{12})^{p_1};n)\prod_{k=1}^{j-1} f(\theta_{2k+1\, 2k+2}^{p_{2k}-p_{2k+1}};n) \right) \left(f(\theta_{1\,2j}^{p_{2j-1}};n) \prod_{k=1}^{j-1} f(\theta_{2k\,2k+1}^{p_{2k}-p_{2k-1}};n) \right)  .
\label{realc}
\eeqa 
By entirely similar arguments it can be shown that 
\beqa 
c_{2j}^{\TT\TT}(\ell,n)&=&\frac{n}{(2j)(2\pi)^{2j} } \sum_{p_1,\ldots, p_{2j-1}=0}^{n-1}\int_{-\infty}^\infty d\theta_1 \cdots \int_{-\infty}^\infty d\theta_{2j} e^{-m\ell \sum_{i=1}^{2j}\cosh\theta_i} \nonumber\\ 
&& \times\,  f(\theta_{1\, 2j}^{p_1};n)f(\theta_{2j-1\, 2j}^{p_1-p_2};n)
 \cdots f(\theta_{23}^{p_{2j-2}-p_{2j-1}};n)f(\theta_{12}^{p_{2j-1}};n).
\label{cc}
\eeqa 
The sums in $p_i$ that enter (\ref{realc})-(\ref{cc}) can be computed exactly for the free Boson and they are given by the formula (\ref{key}) in the appendix. This will allow us to easily analyse the short-distance behaviour of correlators, with the help of formulae (\ref{del})-(\ref{vev}). Let us consider each two-point function separately. 

\section{Leading Short-Distance Behaviours: Extracting the $\log\ell$ Term}

\subsection{The two-Point Function $\bra \TT(0)\tilde{\TT}(\ell)\ket_n$}

Many of the computations in this section are entirely analogous to parts of \cite{nexttonext} where the free Fermion was considered. However, in \cite{nexttonext} some of the computations were only presented in an appendix with limited detail thus we believe it instructive to revisit the steps involved.

Consider the expression (\ref{realc}) and employ the formula (\ref{key}) to perform the sum over the $p_i's$. According to (\ref{key}) the resulting function will depend on the sum of all rapidity dependencies of the functions involved, that is
\beq 
\theta_{12}-\theta_{23}+\theta_{34}+\cdots + \theta_{1\,2j}=-2\sum_{p=1}^j (-1)^p \theta_p=:\theta.
\eeq 
It is convenient to change variables as
\beq 
\theta_{p\, p+1}=x_p \quad \text{for} \quad p=1,\ldots, 2j-1 \quad \text{and} \quad \theta_{2j}=x_{2j}, \label{var1}
\eeq 
we have also that
\beq 
\theta_i=\sum_{p=i}^{2j} x_p \quad \text{for} \quad i=1,\ldots, 2j-1 \quad \text{and} \quad \theta_{2j}=x_{2j}, \label{var2}
\eeq 
so that we may obtain the equivalent of (\ref{genc22}) in terms on the new variables $x_i$ and obtain (\ref{del}) and (\ref{vev}) by integrating over the variable $x_{2j}$. Interestingly, under this change of variables, the sum
\beq 
\theta=2\sum_{p=1}^j x_{2p-1},
\eeq 
which involves only the odd-indexed variables and the difference $\theta_{1\,2j}=\sum_{p=1}^{2j-1} x_p$. This means that the leading small $\ell$ contribution to the function (\ref{realc}) after changing variables and integrating $x_{2j}$ takes the expected form $-x_{\TT\tilde{\TT}} \log \ell$ 
with
\beqa 
x_{\TT\tilde{\TT}} &=& \sum_{j=1}^\infty \frac{2 i n }{j(4\pi)^{2j} }\int_{-\infty}^\infty dx_1 \cdots \int_{-\infty}^\infty dx_{2j-1} \, \,\frac{\mathcal{F}_j(\sum_{p=1}^{j} x_{2p-1},n) \sinh(\sum_{p=1}^{j} x_{2p-1})}{\cosh \frac{\sum_{p=1}^{2j-1} x_{p}}{2}\prod_{i=1}^{2j-1}  \cosh \frac{x_p}{2}}, \label{4666}
\eeqa 
where
\beq 
\mathcal{F}_j(x,n)=\sum_{p=1}^j (-1)^p \left(\begin{array}{c}
2j-1\\
j-p
\end{array}\right)\left[f(2x+(2p-1)i\pi;n)-f(2x-(2p-1)i\pi;n) \right]. \label{curf}
\eeq 
The integral above may be factorised into two functions depending only on even- and odd-indexed variables, respectively.  This may be achieved by introducing the new variable $y=\sum_{p=1}^{j} x_{2p-1}$ (and eliminating the variable $x_{2j-1}$). In terms of this variable we may rewrite some of the $\cosh$ functions in the denominator as follows:
\beq 
\cosh\frac{\sum_{p=1}^{2j-1} x_{p}}{2}= \cosh\left(\frac{y+\sum_{p=1}^{j-1} x_{2p}}{2}\right),
\eeq 
\beq 
\cosh\frac{x_{2j-1}}{2}=\cosh\left(\frac{y-\sum_{p=1}^{j-1} x_{2p-1}}{2}\right).
\eeq 
With this change of variables we find that the integral (\ref{4666}) becomes
\beqa 
x_{\TT\tilde{\TT}} &=& \sum_{j=1}^\infty \frac{2i n }{j(4\pi)^{2j} }\int_{-\infty}^\infty dx_1 \cdots \int_{-\infty}^\infty dx_{2j-2}\int_{-\infty}^\infty dy \, \,\mathcal{F}_j(y,n) \sinh y\nonumber\\
&\times&{\left[\text{sech}\left(\frac{y+\sum_{p=1}^{j-1} x_{2p}}{2}\right)\prod_{p=1}^{j-1}  \text{sech}\frac{x_{2p}}{2}\right]\left[\text{sech}\left(\frac{y-\sum_{p=1}^{j-1} x_{2p-1}}{2}\right)\prod_{p=1}^{j-1} \text{sech} \frac{x_{2p-1}}{2}\right]}. \label{466}
\eeqa 
It was shown in \cite{nexttonext} that these integrals can be performed exactly giving
\beqa 
G_j(y)&=& \int_{-\infty}^\infty dx_1 \cdots \int_{-\infty}^\infty dx_{j-1} \text{sech}\left(\frac{\pm y+\sum_{p=1}^{j-1} x_{p}}{2}\right)\prod_{p=1}^{j-1} \text{sech}\frac{x_{p}}{2} \label{gfun1}\\ 
&=&\int_{-\infty}^\infty da\, \frac{(2\pi)^{j-1} e^{i a y}}{\cosh^j \pi a}\label{gfun2}\\
&=& \frac{(2\pi)^{j-1}}{(j-1)!} \left\{\begin{array}{cc}
\frac{y}{\pi} \text{cosech}\frac{y}{2} \prod_{p=1}^{\frac{j}{2}-1}(\frac{y^2}{\pi^2}+(2p)^2) & \text{for}\quad j \quad \text{even}\\
\text{sech}\frac{y}{2} \prod_{p=1}^{\frac{j-1}{2}}(\frac{y^2}{\pi^2}+(2p-1)^2) & \text{for} \quad j \quad \text{odd}.
\end{array}\right.\label{gfun}
\eeqa 
Thus, the sum (\ref{466}) may be written simply as
\beq 
x_{\TT\tilde{\TT}}= \sum_{j=1}^\infty\frac{2i n }{j(4\pi)^{2j}}
\int_{-\infty}^\infty  dy \, \, \,\mathcal{F}_j(y,n) G_j(y)^2\sinh y. \label{ul}
\eeq 
Note that the integral representation (\ref{gfun1}) only strictly makes sense for $j>1$, although the formulae (\ref{gfun2}) and (\ref{gfun}) are valid for $j\geq 1$ and indeed reproduce the original integral (\ref{4666}) for $j=1$ and $G_1(y)=\text{sech}\frac{y}{2}$. Although (\ref{gfun2}) and (\ref{gfun}) were already used in \cite{nexttonext} it is worth briefly recalling how they follow from (\ref{gfun1}) and from each other. Equation (\ref{gfun2}) can be easily derived by computing the Fourier transform in the variable $y$ of $G_j(y)$ from (\ref{gfun1}). Although (\ref{gfun1}) is a complicated expression, by Fourier transforming in $y$ and then changing variables to $\pm y\rightarrow \pm y-\sum_{p=1}^{j-1} x_p$ all $j$ integrals readily factorize into Fourier transforms of the same function and one obtains 
\beq 
\int_{-\infty}^\infty dy \, G_j(y) e^{iy\omega} = (2\pi)^{j}\text{sech}^{j}(\pi\omega),
\eeq 
from where (\ref{gfun2}) directly follows. This representation can then be employed recursively to obtain the closed formulae (\ref{gfun}). Remarkably the computation of $2j-1$ integrals in formula (\ref{4666}) is then reduced to computing a single integral, which may be easily done numerically.  

Although each contribution to the sum (\ref{ul}) is just an integral of a simple function, it turns out that the sum itself is very slowly convergent for the massive free Boson. However, at least for small integer values of $n$ it is possible to perform the sum very precisely. This is also helped by the fact that the function (\ref{curf}) takes particularly simple forms for $n=2,3,4$ and 6
\beqa 
i\mathcal{F}_j(y,2)\sinh y&=&2^{2(j-1)}, \label{42}\\
i\mathcal{F}_j(y,3)\sinh y&=&3^{j-1}\cosh\frac{y}{3},\\
i\mathcal{F}_j(y,4) \sinh y&=& 2^{j-2}\left(2^{j-1}+\cosh\frac{y}{2}\right),\\
i\mathcal{F}_j(y,6) \sinh y &=&\frac{1}{6}\left(2^{2j-1}+3^j \cosh\frac{y}{3}+\cosh \frac{2y}{3}\right).
\eeqa 
Because of these simple, closed expressions we were able to evaluate the sum (\ref{ul}) up to $j=2000$ giving the results reported in Table~\ref{ta1}.
\begin{table}[h!]
\begin{center}
\begin{tabular}{|l|c|c|c|c|}\hline
  $n$   & 2  &  3 & 4 & 6\\ \hline
  $4\Delta_n$ &$\frac{1}{4}=0.25$ &$\frac{4}{9}=0.444$& $\frac{5}{8}=0.625$& $\frac{35}{36}=0.972$\\ \hline
 $x_{\TT\tilde{\TT}}$ & $0.246$ & $0.438$& $0.608$ & $0.953$\\ \hline
\end{tabular}
\caption{Numerical evaluation of the sum of (\ref{ul}) for $n$ integer with truncation at $j=2000$. The agreement with the predicted values $4\Delta_n$ (as given by (\ref{log1})) is very good even though the sum (\ref{ul}) is very slowly convergent.}
\label{ta1}
\end{center}
\end{table}
In conclusion, the formula (\ref{ul}) reproduces the value $4\Delta_n$ for $n$ integer with great precision (for the data in Table~\ref{ta1} the error remains below 2\%). However, as discussed in \cite{nexttonext}, when $n$ is non-integer, the integral (\ref{ul}) requires a non-trivial analytic continuation. In that case, additional terms need to be added to $x_{\TT\tilde{\TT}}$ which account for the residues of the poles of $\mathcal{F}_j(y,n)$ that cross the real axis as $n \rightarrow 1^+$. The summand in the function $\mathcal{F}_j(y,n)$ has kinematic poles at
\beq 
2y \pm (2p-1)i\pi = (2k n + 1) i \pi \quad \text{and }\quad 2y \pm (2p-1)i\pi = (2k n - 1) i \pi \quad \text{for} \quad k\in \mathbb{Z}.
\eeq 
This poles are due to the presence of kinematic poles of the two-particle form factor (\ref{2pff}) at $\theta=i\pi$ and $\theta=i\pi(2n-1)$, together with its periodicity property $f(\theta;n)=f(-\theta+ 2\pi i n;n)$. This gives rise to four families of poles
\beqa 
y_1&=&(kn+1-p)i\pi, \qquad y_2=(kn-p)i\pi,\qquad k \in \mathbb{Z}\\
y_3&=& (kn-1+p)i\pi,\qquad y_4=(kn+p)i\pi, \qquad k \in \mathbb{Z}, \label{genpoles}
\eeqa 
with corresponding residues of the function inside the sum (\ref{ul})  given by:
\beqa 
R_{1}(j,p,k,n)&=&-\frac{n}{j(4\pi)^{2j}} \left(\begin{array}{c}
2j-1\\
j-p
\end{array}\right)\sinh(i\pi k n) G_j^2((nk-p+1)i\pi),\\
R_{2}(j,p,k,n)&=&-\frac{n}{j(4\pi)^{2j}} \left(\begin{array}{c}
2j-1\\
j-p
\end{array}\right)\sinh(i\pi k n) G_j^2((nk-p)i\pi),\\
R_{3}(j,p,k,n)&=&\frac{n}{j(4\pi)^{2j}} \left(\begin{array}{c}
2j-1\\
j-p
\end{array}\right)\sinh(i\pi k n) G_j^2((nk+p-1)i\pi),\\
R_{4}(j,p,k,n)&=&\frac{n}{j(4\pi)^{2j}} \left(\begin{array}{c}
2j-1\\
j-p
\end{array}\right)\sinh(i\pi k n) G_j^2((nk+p)i\pi).
\eeqa 
Note that all these residues are zero for $n$ integer (due to the presence of the $\sinh(i\pi k n)$ function) so that they only contribute for non-integer $n$. Once we have understood the pole structure of the integrand (\ref{ul}) we must then investigate which of these poles cross the real line in the limit $n\rightarrow 1^+$. This is relatively intricate as the position of each pole depends on $n$, $k$, $j$ and $p$. To ease understanding  
let us consider a simple case as an example: $n=\frac{3}{2}$ and $j=2$ in the sum (\ref{ul}). We know that $4\Delta_{\frac{3}{2}}=0.14$. If we simply evaluate (\ref{ul}) with as much precision as possible we obtain the value $0.0736$ which strongly disagrees with the CFT formula. Moreover this disagreement cannot be entirely explained simply by the truncation of the sum (\ref{ul}). This disagreement is in fact due to the presence of poles of the function $\mathcal{F}_2(y,3/2)$ in (\ref{ul}) which cross the integration line (e.g.~the real axis) as $n$ approaches the value $3/2$.
If we now consider the generic poles (\ref{genpoles}) and the definition (\ref{curf}) we see that for $j=2$ we can only have $p=1,2$. For $p=1$ the four families of poles labeled by the integer $k$ are:
\beqa 
y_1&=&ikn\pi, \qquad y_2=(kn-1)i\pi,\qquad k \in \mathbb{Z}\\
y_3&=& ikn\pi,\qquad y_4=(kn+1)i\pi, \qquad k \in \mathbb{Z}. \label{kpoles}
\eeqa 
Note that the poles at $ikn\pi$ are not double, but arise as single poles of both summands in the function (\ref{curf}). It is clear that these poles are always above the real line (for $k>0$) or below the real line (for $k<0$), that is they never cross the real line, even if $n$ is small. Similarly the poles at $(kn\pm 1)i\pi$ remain above the real line whenever $k>0$ or below the real line if $k<0$ as $n$ approaches $\frac{3}{2}$. Consider now the poles corresponding to $p=2$. We now again have the following four families:
\beqa 
y_1&=&i(kn-1)\pi, \qquad y_2=(kn-2)i\pi,\qquad k \in \mathbb{Z}\\
y_3&=& i(kn+1)\pi,\qquad y_4=(kn+2)i\pi, \qquad k \in \mathbb{Z}. \label{kpoles2}
\eeqa 
We have already seen above that the poles $y_1$ and $y_3$ never cross the real line, so we may at most have some contributions from $y_2$ and $y_4$.
For $k>0$ and $n$ positive and large both families of poles are above the real line. However, for $n=\frac{3}{2}$ we see that the pole $(kn-2)i\pi$ crosses the real line for $k=1$. Similarly, for $k<0$ and $n$ positive and large all poles are in the lower half plane but the pole $(kn+2)i\pi$ crosses the real line for $\frac{3}{2}$ and $k=-1$. 

In summary, there are two poles for $j=p=2$ located at $\pm \frac{i\pi}{2}$ that cross the real line as $n\rightarrow \frac{3}{2}$. The corresponding residue contributions are
\beqa  
2\pi i (R_2(2,2,1,3/2)- R_4(2,2,-1,3/2))&=&-\frac{3i}{2^8\pi^{3}} \sinh\frac{3i\pi}{2} \left(G_2^2(-\frac{i\pi}{2})+G_2^2(\frac{i\pi}{2})\right)\nonumber\\
&=&-\frac{3}{2^6\pi}=-0.0149208. \label{exam32}
\eeqa  
Therefore, the addition of the residua of these two poles improves the estimate of the conformal dimension from 0.0736 to the value 0.0885 (note that the formula (\ref{ul}) gives -$4\Delta_n$, hence the minus sign of (\ref{exam32})). Similarly, the addition of poles for higher values of $j$ will bring this value ever closer to $4\Delta_{\frac{3}{2}}=0.14$ as shown in Fig~2.

In the general $n$ case, in order to fully identify those poles that will cross the real line we find once more four cases:
\beqa  
y_1: kn+1-p <0 \quad &\Rightarrow& \quad 1 \leq k < \frac{p-1}{n},\nonumber\\
y_2: kn-p<0 \quad &\Rightarrow& \quad 1 \leq k <\frac{p}{n},\nonumber\\
 y_3: kn-1+p <0 \quad &\Rightarrow & \quad -\frac{p-1}{n} < k \leq -1,\nonumber\\
 y_4: kn+p<0 \quad & \Rightarrow & \quad -\frac{p}{n} < k \leq -1,\label{ranges}
\eeqa  
This gives the analytically continued values
\beqa 
\tilde{x}_{\TT\tilde{\TT}}&=&x_{\TT\tilde{\TT}}-\sum_{j=1}^\infty\sum_{p=1}^j \sum_{k=1}^{[\frac{p-1}{n}]-q_1} \frac{i n}{j (4\pi)^{2j-1}} \left(\begin{array}{c}
2j-1\\
j-p
\end{array}\right) \sinh\left(i\pi n k\right) G_j^2\left(\left(nk-p+1\right)i\pi\right)\nonumber\\
&&\qquad -\sum_{j=1}^\infty\sum_{p=1}^j \sum_{k=1}^{[\frac{p}{n}]-q_2} \frac{i n }{j (4\pi)^{2j-1}} \left(\begin{array}{c}
2j-1\\
j-p
\end{array}\right) \sinh\left(i\pi n k\right) G_j^2\left(\left(nk-p\right)i\pi\right). \label{anul}
\eeqa 
The shifts $q_1, q_2$ take the value 1 when $n[\frac{p-1}{n}]=p-1$ and $n[\frac{p}{n}]=p$, respectively and are zero otherwise. Here the symbol $[.]$ represents the integer part.
\begin{figure}[h!]
\begin{center}
\includegraphics[width=10cm]{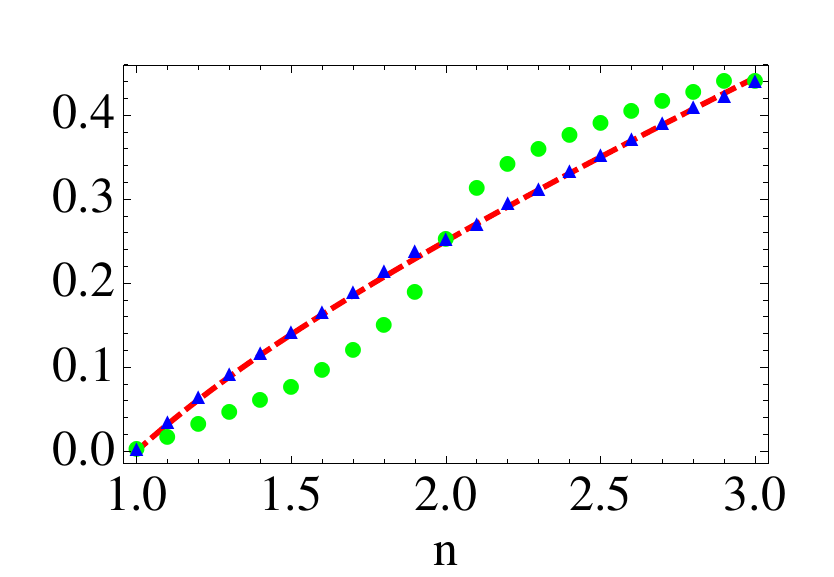}
\caption{The dashed line is the function $4\Delta_n=\frac{1}{6}\left(n-\frac{1}{n}\right)$. The circles are the values of $x_{\TT\tilde{\TT}}$ as given by (\ref{ul}) and the triangles are the values of $\tilde{x}_{\TT\tilde{\TT}}$ as given by (\ref{anul}).  Clearly the extra poles included in (\ref{anul}) give a very sizable contribution for non-integer values of $n$. }
\label{atlast}
\end{center}
\end{figure}

To conclude this section, we note once more that both the sequence (\ref{ul}) and (\ref{anul}) are very slowly convergent. Even after the inclusion of 2000 terms in Table 1 agreement with analytical results is not perfect. 
The values depicted in Fig.~\ref{atlast} show almost perfect agreement with the analytical result but only because we have managed to sum (\ref{ul}) and (\ref{anul}) almost exactly. We achieved this by first truncating each sum up to $j=150$ and then carrying out a linear fit of the logarithm of individual terms from $j=20$ to $j=150$ against $\log j$. Such fit is extremely precise and we could then use it to carry out the rest of the sum (from $j=151$ to $\infty$). This latter sum turns out to still give an important contribution to the final value (around 8\%). 

This is rather surprising given that a previous investigation of the free Fermion, where very similar expressions emerge leads to rapidly convergent sequences and very accurate predictions, as shown in \cite{nexttonext}. Despite this observation, the numerical results depicted in Fig.~\ref{atlast} provide strong evidence for (\ref{anul}) representing the correct analytic continuation to $n$ non-integer. Despite the slight disagreement with the analytical formula, it is clear from Fig.~\ref{atlast} that (\ref{ul}) either under- or overstimates the value of $4\Delta_n$ if $n$ is non-integer and that it has oscillations which are smoothed out by the addition of the residues associated with the poles (\ref{ranges}) which cross the real line as $n$ approaches 1. 

As we will see, convergence issues appear to be a typical feature of the massive free Boson theory and will feature again when we compute other physical quantities. We will discuss their possible origin in sections 6, 8 and Appendix B. 
 
\subsection{The two-Point Function $\bra \TT(0){\TT}(\ell)\ket_n$}
Once again we use the formula (\ref{key}) to carry out the sum over the indices $p_i$ in  (\ref{cc}). The result depends on the sum of all rapidity dependencies entering the two particle form factors $f(\theta;n)$ in the sums. In this case this leads to a remarkable simplification as
\beq 
\theta_{12}+\theta_{23}+ \cdots+\theta_{2\ell-1\, 2\ell}+\theta_{2j\,\, 1}=0,
\eeq 
by construction. This means that the value of the sum in (\ref{cc}) is given by the particular limiting case of (\ref{key}), which after analytic continuation in $n$ is given by (\ref{magic}). Thus we have that
\beqa 
c_{2j}^{\TT\TT}(\ell,n)&=&\frac{n \,h(j,n)}{2j(2\pi)^{2j}}
\int_{-\infty}^\infty d\theta_1 \cdots \int_{-\infty}^\infty d\theta_{2j} \, \prod_{i=1}^{2j} \text{sech}\frac{\theta_{i \, i+1}}{2}  e^{-m\ell \sum_{i=1}^{2j}\cosh\theta_i},
\label{ccx}
\eeqa 
or, after introducing the variables $x_i$ defined earlier (\ref{var1})-(\ref{var2}), integrating over the variable $x_{2j}$ and expanding the Bessel function as in (\ref{genc222}) we obtain
\beq 
x_{\TT\TT}=\sum_{j=1}^\infty \frac{n \,h(j,n)\,}{j(2\pi)^{2j}}\int_{-\infty}^\infty d x_1 \cdots \int_{-\infty}^\infty d x_{2j-1} \,\text{sech} \left(\frac{\sum_{p=1}^{2j-1} x_p}{2}  \right)\prod_{p=1}^{2j-1} \text{sech}\frac{x_p}{2}.
\label{xtt}
\eeq 
The integrals above are special cases of formula (\ref{gfun}) which allows for their direct evaluation. Note that they are entirely independent of the value of $n$ which only enters through the function $n h(j,n)$. It is easy to show that (this is just a special case of (\ref{gfun}))
\beqa 
\int_{-\infty}^\infty d x_1 \cdots \int_{-\infty}^\infty d x_{2j-1} \,\text{sech} \left(\frac{\sum_{p=1}^{2j-1} x_p}{2}  \right)\prod_{p=1}^{2j-1} \text{sech}\frac{x_p}{2}
= \frac{(4\pi)^{2j-1}}{(2j-1)!}\frac{1}{\pi}((j-1)!)^2. \label{vj}
\eeqa 
Substituting (\ref{vj}) into (\ref{xtt}) we obtain the sum
\beq 
x_{\TT\TT}=\frac{n}{4\pi^2}\sum_{j=1}^\infty \frac{2^{2j} h(j,n) ((j-1)!)^2}{j (2j-1)!}.\label{cn}
\eeq 
Employing the definition of $h^{e}(j,n)$ given in (\ref{heven}) we have that
\beqa 
x_{\TT\TT}^e&=&\frac{n}{2\pi^2}\sum_{j=1}^\infty \frac{((j-1)!)^2}{j (2j-1)!} \left[\left(\begin{array}{c}
2j-1\\
j-1
\end{array}\right) + \sum_{p=1}^{[\frac{j}{n}]} 
\left(\begin{array}{c}
2j\\
j-p n
\end{array}\right)\right]\nonumber\\
&=&\frac{n}{2\pi^2}\sum_{j=1}^\infty\frac{1}{j^2}+\frac{n}{2\pi^2}\sum_{j=1}^\infty\sum_{p=1}^{[\frac{j}{n}]} \frac{((j-1)!)^2}{j (2j-1)!}
\left(\begin{array}{c}
2j\\
j-p n
\end{array}\right)\nonumber\\
&=& \frac{n}{12}+\frac{n}{2\pi^2}\sum_{p=1}^\infty \sum_{j=n p}^{\infty} \frac{((j-1)!)^2}{j (2j-1)!} 
\left(\begin{array}{c}
2j\\
j-p n
\end{array}\right)=\frac{n}{12}+\frac{1}{6n},\label{sum}
\eeqa 
and similarly for $n$ odd. We find
\beq 
x_{\TT\TT}^{e}=\frac{n}{12}+\frac{1}{6n} \quad \text{and} \quad x_{\TT\TT}^o=\frac{n}{12}-\frac{1}{12n}. \label{xx}
\eeq 
Here the $e$ and $o$ superindices indicate analytic continuations of $x_{\TT\TT}$ from $n$ even and odd, respectively. The values above are exactly those predicted by CFT as seen in the leading contributions to (\ref{log2}). The expected results are obtained for generic $n$, thus showing that the functions $h^{e,o}(j,n)$ indeed capture the correct analytic continuation from $n$ integer and even or odd to $n$ real and positive. In particular, by setting $n=1$ in $x_{\TT\TT}^e$ we recover the value $\frac{1}{4}$ in line with CFT predictions for the logarithmic negativity \cite{negativity1,negativity2,ourneg}.  It is worth emphasizing that the results (\ref{xx}) follow from exactly re-summing all the infinitely many terms resulting from a form factor expansion, something that can rarely be achieved for any QFT local fields.  

\section{Next-to-Leading Order Short-Distance Behaviours: Expectation Values and Structure Constants}
 \subsection{The two-Point Function $\bra \TT(0) \tilde{\TT}(\ell)\ket_n$}
\begin{figure}[h!]
\begin{center}
\includegraphics[width=7.5cm]{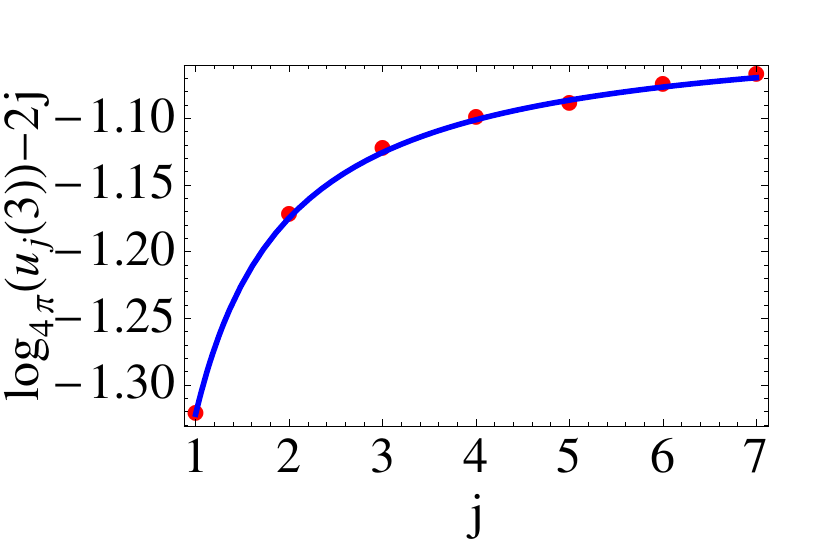}
\includegraphics[width=7.5cm]{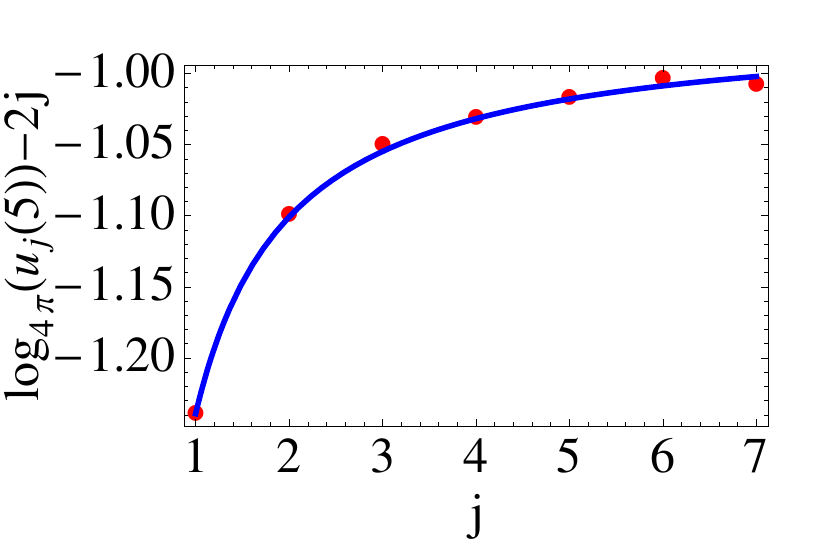}
\caption{The red dots show the numerical values of $\log_{4\pi}(u_j(n))-2j$ for $n=3$ and $n=5$ and $j\leq 7$ evaluated using a Monte Carlo approach. The blue lines are fits of the form $\alpha+\frac{\beta}{j}$.}
\label{upic}
\end{center}
\end{figure}

 Consider the expression (\ref{log1}) together with (\ref{o1o2}) and (\ref{vev}). We may now evaluate $K_{\TT\tilde{\TT}}=2\log\bra \TT \ket_n$ by employing (\ref{vev}) and the results of the previous section. In particular, we will use the variables (\ref{var1})-(\ref{var2}) in terms of which we obtain
 \beq
 K_{\TT\tilde{\TT}}
:=2 \log \bra\TT \ket_n= x_{\TT\tilde{\TT}}(\log\frac{m}{2}+\gamma)+\sum_{j=1}^{\infty}\frac{2 n \, u_j(n)}{j(4\pi)^{2j}}, \label{kttt}
 \eeq
 where
 \beq 
u_j(n)=\int_{-\infty}^\infty dx_1 \cdots \int_{-\infty}^\infty dx_{2j-1} \, \,\frac{i\mathcal{F}_j(\sum_{p=1}^{j} x_{2p-1},n) \sinh(\sum_{p=1}^{j} x_{2p-1})}{\cosh \frac{\sum_{p=1}^{2j-1} x_{p}}{2}\prod_{p=1}^{2j-1}  \cosh \frac{x_p}{2}}\log d_j,\label{uln}
\eeq 
and $d_j$ are the functions defined in (\ref{dj}) now expressed in terms of the variables $x_1, \ldots, x_{2j-1}$ as 
 \beq 
d_j^2={\left(\sum_{i=1}^{2j-1} \cosh\left(\sum_{p=i}^{2j-1} x_p\right)+1\right)^2-\left(\sum_{i=1}^{2j-1} \sinh\left(\sum_{p=i}^{2j-1} x_p\right)\right)^2}. \label{dd}
 \eeq 
The integrals $u_j(n)$ can be computed by means of Monte Carlo integration procedures (except for $n=1$ where $u_j(1)=0$ directly from the definition) leading for instance to the values depicted in Fig.~\ref{upic}.
The numerical values obtained for $u_j(n)$ are in all cases best fitted by functions of the form
\beq 
u_j^{\text{fit}}(n)=(4\pi)^{2j+a_n+\frac{b_n}{j}}. \label{fituj}
\eeq 
In principle we could use these fits to evaluate the sum (\ref{kttt}) up to large values of $j$. However, it is clear from the fits (\ref{fituj}) that
\beq 
\lim_{j\rightarrow \infty} \frac{u_j^{\text{fit}}(n)}{(4\pi)^{2j}}=(4\pi)^{a_n},
\eeq 
which means that the sum (\ref{kttt}) is divergent, even if each individual integral $u_j(n)$ takes a finite value. This is an a priori surprising result which needs to be physically understood. A discussion and interpretation of this result will be presented in section~6. We will show that despite the sum (\ref{kttt}) being divergent, we may still extract useful information from it.  

 \subsection{The two-Point Function $\bra \TT(0) {\TT}(\ell)\ket_n$}
 Let us go back to formulae (\ref{log2}), (\ref{o1o2}) and (\ref{vev}) and let us examine the next to leading order correction to (\ref{log2}), that is the ratios of expectation values and three-point couplings of the twist field given in (\ref{log2}). According to (\ref{vev}) and employing once more the variables (\ref{var1})-(\ref{var2}) we can write 
 \beq
 K_{\TT{\TT}}
= x_{\TT{\TT}}(\log\frac{m}{2}+\gamma)+\sum_{j=1}^{\infty}\frac{n \,h(j,n) v_j}{j(2\pi)^{2j}}. \label{ktt}
 \eeq
 where
 \beq 
 v_j=\int_{-\infty}^\infty d x_1 \cdots \int_{-\infty}^\infty d x_{2j-1} \,\text{sech} \left(\frac{\sum_{p=1}^{2j-1} x_p}{2}  \right)\prod_{p=1}^{2j-1} \text{sech}\frac{x_p}{2} \log d_j, \label{vej}
 \eeq 
 and $d_j$ are the functions (\ref{dd}).
 
\begin{figure}[h!]
\begin{center}
\includegraphics[width=9cm]{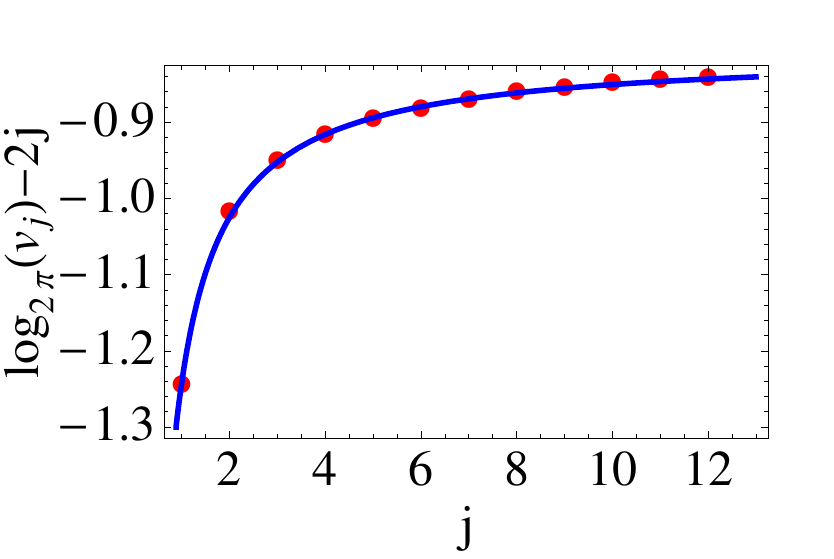}
\caption{The red dots show the numerical values of $\log_{2\pi}(v_j)-2j$ evaluated using a Monte Carlo approach. The blue line is the function $\log_{2\pi}\frac{v_j^{\text{fit}}}{(2\pi)^{2j}}=-0.806996 - \frac{0.436331}{j}$. The fit is extremely good indicating that for $j$ large the $\log_{2\pi}(v_j)$ is linear in $j$.}
\label{ama}
\end{center}
\end{figure}
A crucial feature of the integrals $v_j$ is that they are $n$-independent. Besides the case $j=1$ where 
\beq 
v_1=u_1(2)=\frac{1}{2}\int_{-\infty}^\infty dx \, \frac{\log(2(1+\cosh x))}{\cosh^2\frac{x}{2}}=4,
 \eeq 
we have not found closed formulae for $j>1$ but we have been able to compute the integrals very precisely through Monte Carlo integration procedures. Fig.~\ref{ama} shows the numerically obtained values of $v_j$ for $j\leq 12$. These values are very precisely fitted by the curve
\beq 
v_j^{\text{fit}}= (2\pi)^{2j-0.806996 -\frac{0.436331}{j}}. \label{fit}
\eeq 
We may now use this fit to extrapolate to larger values of $j$ (rather than carrying out the integrals). In this way, we will be able to perform the sum (\ref{ktt}) up to very large values of $j$. For $n$ odd, we obtain the values reported in Table~\ref{ratiodd}.  

\begin{figure}[h!]
\begin{center}
\includegraphics[width=9cm]{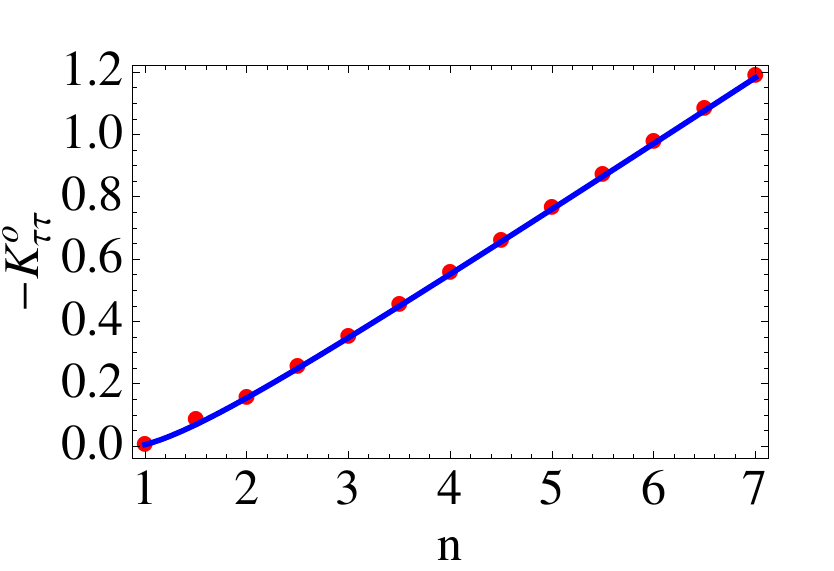}
\caption{The red dots show the numerical values of $-K_{\TT\TT}^o$ evaluated using formula (\ref{ktt}) with $h^o(j,n)$ for various (non-integer and integer) values of $n$ and summing as many terms as needed to ensure convergence. The blue line is the function $-0.34+ \frac{0.13}{n} + 
 0.215 n$. The fit is extremely good indicating that the ratio of $\mathcal{C}_{\TT\TT}^{\TT^2}$ and $\bra \TT\ket_n$ decays exponentially for $n$ odd. Here, as before, we have set the mass scale $m=1$.}
\label{amar}
\end{center}
\end{figure}

\begin{figure}[h!]
\begin{center}
\includegraphics[width=10cm]{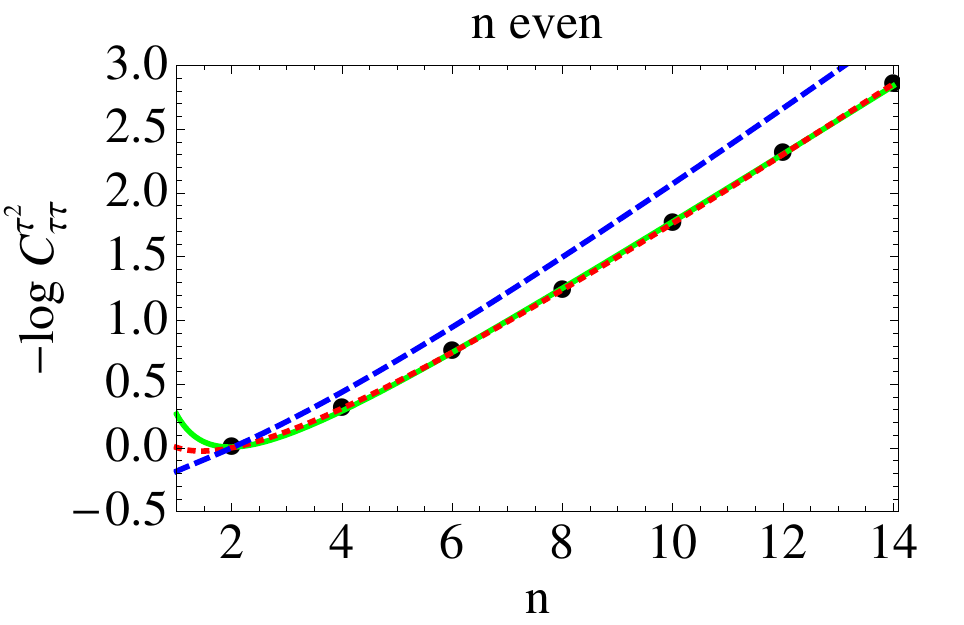}
\caption{The logarithm of the structure constant $\mathcal{C}_{\TT \TT}^{\TT^2}$ for even values of $n$ (dots). The solid and dotted lines provide two possible fits of the points obtained. The solid (green) curve is the function $-\log(\mathcal{C}_1(n))=-1.074+ \frac{1.064}{n} + 0.274 n$ and the dotted (red) curve is the function $-\log(\mathcal{C}_2(n))=-0.308 + 0.311 n - 
 0.456 \log n $. As we can see both are extremely good for the points we have and yet their curvatures around $n=1$ are quite different. The dashed blue line represents an analytical prediction given in \cite{negativity2} (see section 6 for discussions on this comparison).}
\label{5}
\end{center}
\end{figure}

From (\ref{log2}) we have that $K^{o}_{\TT\TT}=-\log\frac{\mathcal{C}_{\TT \TT}^{\TT^2}}{\langle \TT \rangle_n}$ so that the formula (\ref{ktt}) provides a prediction for a ratio of two universal QFT quantities. 
\begin{table}[h!]
\begin{center}
\begin{tabular}{|l|c|c|c|c|c|c|c|c|c|}\hline
  $n$   & 3  &  5 & 7 & 9 & 11 & 13 & 15&17&19\\ \hline
  $-K_{\TT\TT}^o$ &  $0.345$  & $0.760$ &$1.183$ &$1.607$ & $2.033$ &$2.459$&$2.885$& $3.311$ & $3.737$\\ \hline
\end{tabular}
\caption{Numerical values of $-K_{\TT\TT}^o=\log\frac{\mathcal{C}_{\TT \TT}^{\TT^2}}{\langle \TT \rangle_n}$ for $n$ odd summing up to as many terms as needed to see convergence. The values are obtained by evaluating the sum (\ref{ktt}) employing the fit (\ref{fit}) and setting the mass scale $m=1$.}
\label{ratiodd}
\end{center}
\end{table}
Further, because the full $n$-dependence is encapsulated by the function $h^o(j,n)$, we may also use the formula (\ref{ktt}) for non-integer or even values of $n$. A graphical representation of the values obtained for $n\leq 7$ is given in Fig.~\ref{amar}. As can be seen, we obtain a smooth function of $n$ which displays linear behaviour for large $n$. In particular it is easy to show that $h^o(1,j)=0$ for all $j$ and therefore we have that 
\beq 
\lim_{n_o\rightarrow 1}\log\frac{\mathcal{C}_{\TT \TT}^{\TT^2}}{\langle \TT \rangle_n}=0.
\eeq 
This result is exactly what we would expect since $\bra \TT \ket_1=1$ and $\Delta_1=0$ (for $n=1$ we only have one replica so the twist field becomes the identity field). Also, for $n$ odd the field $\TT^2=\TT$ and so $\mathcal{C}_{\TT\TT}^{\TT^2}=\mathcal{C}_{\TT\TT}^\TT$. For $n=1$ this is the structure constant associated with the identity field which should be also 1. 
We may attempt now to perform the same sum (\ref{ktt}) employing the function $h^e(j,n)$ defined in (\ref{heven}). This should provide an analytic continuation from $n$ even of the function 
\beq 
K^e_{\TT\TT}=-\log\frac{\langle \TT \rangle_{\frac{n}{2}}^2\mathcal{C}_{\TT \TT}^{\TT^2}}{\langle \TT \rangle_n^2}. \label{ke}
\eeq 
Unfortunately, the sum (\ref{ktt}) (similar to (\ref{kttt})) is divergent for $n$ even. The difference with respect to the $n$ odd case is due to the asymptotic properties
\beq 
\lim_{j\rightarrow \infty} h^e(j,n)=\frac{1}{n} \quad \text{and} \quad \lim_{j\rightarrow \infty} h^o(j,n)=0.
\eeq 
It is however possible to evaluate $\mathcal{C}_{\TT\TT}^{\TT^2}$ by subtracting the divergent sum (\ref{kttt}) from (\ref{ktt}) in such a way as to remove all dependence on the expectation values. In other words, we may compute 
\beqa 
 \log \mathcal{C}_{\TT \TT}^{\TT^2}= -n \sum_{j=1}^\infty \left(
 \frac{h^e(j,n)v_j}{j(2\pi)^{2j}}+\frac{u_j(\frac{n}{2})-2 u_j(n)}{j (4\pi)^{2j}}\right). \label{cancel}
 \eeqa 
 In particular, for $n=2$ we can employ the fact that $u_j(1)=0$ and $u_j(2)=2^{2(j-1)} v_j$ (this is due to the equality (\ref{42})) to find
 \beqa 
 \lim_{n_e\rightarrow 2}\log \mathcal{C}_{\TT \TT}^{\TT^2} = \sum_{j=1}^\infty \frac{(1-2 h^e(j,2))v_j}{j (2\pi)^{2j}}=0. \label{goodsum}
 \eeqa 
The result above follows trivially from the property $h^e(j,2)=\frac{1}{2}$ $\forall$ $j$ and gives a neat example of how the difference of two divergent series may produce a convergent one. The sum above is identically zero (irrespective of the values of $v_j$), giving us the exact result
 \beq 
 \lim_{n_e\rightarrow 2}\mathcal{C}_{\TT \TT}^{\TT^2}=1.
 \eeq 
This result is in agreement with what we expect from CFT considerations. It was first argued in \cite{MB} that the field $\TT^2$ is nothing but the identity field for $n=2$ and so the result follows from the CFT normalization of correlators. For other values of $n$ we rely on the numerical fits $v_j^{\text{fit}}$ and $u_j^{\text{fit}}(n)$ which are of course not exact. However, within the error of these fits we have been able to show that the sum (\ref{cancel}) is indeed convergent.  Fig.~\ref{5} shows our results for several even values of $n$.
The solid (green) and dotted (black) lines presented in Fig.~\ref{5} are fits which provide a numerical analytic continuation from $n$ even to $n$ real and positive. In particular our numerical values for $\mathcal{C}_{\TT\TT}^{\TT^2}$ are very well fitted by either $\mathcal{C}_1(n)=e^{1.074-\frac{1.064}{n}-0.274n}$ or $\mathcal{C}_2(n)=e^{0.308-0.311n+0.456\log n}$ and allow us to obtain the following values
\beq 
\mathcal{C}_{1}(1)=0.77 \quad \text{and} \quad 
 \mathcal{C}_{2}(1)=1.0.
\label{cttt2}
\eeq 
These results can be compared to an analytic prediction in \cite{negativity2} where the value of the structure constant was computed for the compactified free Boson in the double limit $n\rightarrow 1$ and $\eta \rightarrow \infty$ where $\eta$ is the compactification radius. The value predicted in \cite{negativity2} is 
\beq 
\lim_{n_e\rightarrow 1^+}\mathcal{C}_{\TT\TT}^{\TT^2}=\frac{A^6}{2^{7/6} e^{1/2}}=1.20184...\label{exact}
\eeq 
where $A=1.2824...$ is Glaisher's constant and in \cite{negativity2} this number was called $P_1^{-1}$. 
This analytical value lies slightly above both values (\ref{cttt2}) and is closest to the value  $\mathcal{C}_2(1)$. This highlights the difficulty of performing reliable analytic continuations based solely on a (small) number of numerically obtained values. The fact that the fit $\mathcal{C}_2(n)$ seems to work best near $n=1$ is natural once we notice that the analytical prediction (dashed line) also has an expansion of the form $a+bn+c\log n$ for large $n$ (see discussion in section 6).

\section{Interpretation of Divergent Series and $\log\log$-Corrections}
In the previous sections we have shown that a form factor approach allows accurate access to leading and next to leading order short distance corrections to twist field two-point functions. Such corrections involve universal quantities which characterize both the massive theory and its conformal counterpart. They are given by expectation values $\bra\TT \ket_n$ and the three-point coupling $\mathcal{C}_{\TT\TT}^{\TT^2}$ of twist fields, both of which are generally very hard to compute analytically, even for free theories. A feature of particular interest is that for the massive free Boson the form factor expansions of 
\beq 
\log \bra \TT \ket_n \qquad \text{and} \qquad   \log \frac{\bra \TT\ket^2_{\frac{n}{2}} \mathcal{C}_{\TT\TT}^{\TT^2}}{\bra\TT\ket_n^2} \quad \text{for n even}
\eeq 
give rise to divergent sums. Our interpretation of such divergences is that they arise from the presence of (unaccounted for) logarithmic divergences of the corresponding correlators. In other words, the formulae (\ref{log1}) and (\ref{log2}) do not capture the true $\ell$-dependence of the correlators and this in turn means that the identification of $K_{\mathcal{O}_1\mathcal{O}_2}$ through formula (\ref{vev}) is not entirely justified. However, remarkably, our functions $K_{\TT\TT}$ and $K_{\TT\tilde{\TT}}$ still capture universal QFT information which is revealed when special divergence-canceling combinations of correlators are evaluated. We propose that (\ref{log1})-(\ref{log2}) should be replaced by 
\beq
\log\left(\frac{\langle \TT(0)\tilde{\TT}(\ell) \rangle_n}{\langle \TT \rangle_n^2}\right)_{m\ell \ll 1}=-4\Delta_n \log \ell-r_1(n)\log (p \log\ell) -2 \log \bra \TT \ket_n. \label{log11}
\eeq 
Similarly 
\beq 
\log\left(\frac{\langle \TT(0){\TT}(\ell) \rangle_n}{\langle \TT \rangle_n^2}\right)_{m\ell \ll 1}=\left\lbrace\begin{array}{cc}
-2\Delta_n \log\ell+\log\frac{\mathcal{C}_{\TT \TT}^{\TT^2}}{\langle \TT \rangle_n}& \text{for n odd}\\
-4(\Delta_n-\Delta_{\frac{n}{2}})\log\ell- r_2(n)\log (p \log\ell) +\log\frac{\langle \TT \rangle_{\frac{n}{2}}^2\mathcal{C}_{\TT \TT}^{\TT^2}}{\langle \TT \rangle_n^2}& \text{for n even}\\
\end{array}\right.\label{log22}
\eeq 
where $r_1(n)$ and $r_2(n)$ are unknown functions and $p$ is a constant. An analytic calculation for the massless free Boson showing the emergence of a $\log(\log \ell)$ correction in (\ref{log11}) will be presented shortly in \cite{Part2}.
Obviously the presence of the constant $p$ is equivalent to a redefinition of $K_{\TT\TT}$ and $K_{\TT\tilde{\TT}}$ and this means that there is naturally a certain ambiguity in the identification of the expectation values and three-point couplings through this approach. 

From our form factor computation, the $n$-dependence of the functions $r_1(n)$ and $r_2(n)$ can be fixed by imposing the cancellation of divergences that we have numerically observed. 
There are three key observations that we may use:
\begin{itemize}
\item[1)] The fact that the sum (\ref{cancel}) is convergent implies that
\beq 
r_1(n)-r_1(\frac{n}{2})=r_2(n).
\eeq 
\item[2)] Another numerical observation which was suggested by preliminary results of 
 \cite{Part2} is that the ratio
 \beq
2\log \frac{\bra \TT \ket_n}{\bra\TT \ket_2^{n-1}}, \label{tn}
\eeq 
also admits a convergent form factor expansion representation even if the expansion of $\bra \TT\ket_n$ itself is divergent. 
 The cancellation of divergences in (\ref{tn}) is equivalent to requiring
\beq 
r_1(n)-(n-1)r_1(2)=0,
\eeq 
that is $r_1(n)=r (n-1)$ with $r$ constant. 
\item[3)] We may even determine the sign of this constant $r$ even if form factors alone cannot fix its value. This is because the expansion (\ref{kttt}) is not just divergent but tends to $+\infty$ (all functions involved in the sum are positive definite). This observation means that whatever the correction to the leading $\log \ell$ term is, its effect should be to reduce its value (note that the factor $K_{\TT\tilde{\TT}}$ appears with a negative sign in the expansion (\ref{log1}). This means that $r>0$.
\end{itemize}
The presence of logarithmic divergences in the correlators of the massive free Boson is not entirely surprising as we are dealing from the beginning with an underlying logarithmic CFT. It was shown in \cite{BCDLR} in complete generality that an additional contribution to the R\'enyi entropy of the form $r\log(\log\ell)$ will always emerge when dealing with logarithmic CFTs \cite{gurarie} (see eq.~(13)). In this context, the coefficient $r$ was shown to be a positive integer, which is related to the algebraic structure of the CFT. 

In the specific case of the non-compactified massless free Boson, additional $\log\log$ divergences of other twist field correlators at criticality are also found when studying the LN of adjacent regions in the compactified free Boson in the limit when the compactification radius $\eta \rightarrow \infty$ \cite{negativity2}  and also when studying the EE of two disconnected regions. The presence of a $\log(\log\ell)$ correction in (\ref{log11}) can in fact be inferred directly from the results of \cite{log} where the four-point function $\bra \TT(-\ell_1)\tilde{\TT}(0)\TT(\ell_2)\tilde{\TT}(\ell_2+\ell_3)\ket_n$ of the compactified massless free Boson was investigated. Combining Eq.~(4) and Eq.~(66) in \cite{log} it was found that for large compactification radius $\eta$ at criticality
\beqa 
\bra \TT(-\ell_1)\tilde{\TT}(0)\TT(\ell_2)\tilde{\TT}(\ell_2+\ell_3)\ket_n=\frac{g(\ell_1,\ell_2,\ell_3)^{4\Delta_n} \eta^{n-1}}{\prod_{k=1}^{n-1}{\,}_2F_1(\frac{k}{n},1-\frac{k}{n},1;x) {\,}_2F_1(\frac{k}{n},1-\frac{k}{n},1;1-x)},
\eeqa 
where $g(\ell_1,\ell_2,\ell_3)$ is a known ratio of lengths and $x=\frac{\ell_1\ell_3}{(\ell_1+\ell_2)(\ell_2+\ell_3)}$ is the usual cross-ratio. It is easy to see that the leading term in the expansion of the functions above about
 $\ell_2=0$ (or $x=1$) is given by
\beq 
\left(\bra \TT(-\ell_1)\tilde{\TT}(0)\TT(\ell_2)\tilde{\TT}(\ell_2+\ell_3)\ket_n\right)_{x \approx 1} =  \frac{(\ell_2(\ell_1+\ell_3))^{-4\Delta_n} \eta^{{n-1}}}{\prod_{k=1}^{n-1} \frac{-\log(1-x)}{\Gamma(\frac{k}{n})\Gamma(1-\frac{k}{n})}}=
 \frac{(\ell_2(\ell_1+\ell_3))^{-4\Delta_n} (2\pi \eta)^{n-1}}{n (-\log(1-x))^{n-1}}. 
\eeq 
Therefore we have that the von Neumann entropy diverges as
\beq 
\lim_{n\rightarrow 1}\frac{ \log(\bra \TT(-\ell_1)\tilde{\TT}(0)\TT(\ell_2)\tilde{\TT}(\ell_2+\ell_3)\ket_n)_{x\approx 1}}{1-n}=\frac{1}{3}\log(\ell_2(\ell_1+\ell_3))+ \log(-\log(1-x))+\cdots \label{xss1}
\eeq 
This suggests that the constant $r=1$ and therefore\footnote{We thank P.~Calabrese and E.~Tonni for discussions and clarification of this point.} 
\beq 
r_1(n)=n-1 \qquad \text{and} \qquad r_2(n)=\frac{n}{2}.
\eeq  
Another correlator of interest was considered in \cite{negativity2}. It was shown that the LN of adjacent regions in the massless (non-compactified) free Boson behaves as
\beq 
\mathcal{E}(\ell_1,0,\ell_2)=\lim_{n_e \rightarrow 1^+}\log(\bra \TT(-\ell_1)\tilde{\TT}^2(0)\TT(\ell_2)\ket_n)=\frac{1}{4}\log y-\frac{1}{2}\log\left(\frac{1}{2}\log y\right)-\log P_1+ O(1)
\eeq 
where $y=\frac{\ell_1 \ell_2}{\ell_1+\ell_2}$ and $P_1$ is the inverse three-point coupling
\beq 
-\log P_1=\log\mathcal{C}_{\TT\tilde{T}^2{\TT}}\equiv \log \mathcal{C}_{\TT\TT}^{\TT^2}, \label{3pt}
\eeq 
(see \cite{MB} for a discussion of various equivalences between three-point couplings of twist fields such as the one used above). Note that once more a $\log\log$ correction is present which appears with the same coefficient as in (\ref{log22}) when the same limit  $n_e \rightarrow 1^+$ is taken.
The identification (\ref{3pt}) once more suffers from the ambiguity of whether or not the term $1/2 \log 2$ which is included in the double logarithm should be identified as part of the three-point coupling. The numerical comparison in Fig.~6 suggests that $-\log P_1$ should indeed be identified with $\log \mathcal{C}_{\TT\TT}^{\TT^2}$ in our set-up. In fact, we can even compare our results to those given in \cite{negativity2} beyond the $n=1$ analytic continuation. A full expression for the constant $P_n$ with $n$ even was derived in \cite{negativity2} and is given by
\beq 
P_n=\frac{2\pi^{(n-3)/2}}{\sqrt{n}}\exp \int_{0}^\infty \frac{dt e^{-t}}{t} \left(\frac{1}{1-e^{-t}}\left(\frac{e^{\frac{t}{2}}-1}{e^{\frac{t}{n}}-1}-\frac{n}{2}\right)-\frac{n-2}{8}\right).\label{paspn}
\eeq 
However, apart from the inherent ambiguity in the definition of the constants $P_n$ emerging from the presence of terms such as $\log\left(\frac{1}{2}\log\ell\right)$ above, there is also another ambiguity emerging from the fact that the computations in \cite{negativity2} are done for a compactified free Boson and depend on the compactification radius through a factor $\eta^{\frac{(n-1)}{2}}$. In \cite{negativity2} it is argued that by taking first the limit $n\rightarrow 1$ and then $\eta \rightarrow \infty$ results for the decompactified free Boson should be obtained, among them the constant $P_1$. However, if we are to compare our numerical values in Fig.~6 to formula (\ref{paspn}) then the presence of the factor  $\eta^{\frac{(n-1)}{2}}$ can play a role. It is of course rather hard to asses how this infinite constant (for $n$ even and $\eta \rightarrow \infty$) affects the definition of $\mathcal{C}_{\TT\TT}^{\TT^2}$. Our benchmark has been to use the fact that $\lim_{n_e\rightarrow 2}\mathcal{C}_{\TT\TT}^{\TT^2}=1$. It turns out that if we identify $\mathcal{C}_{\TT\TT}^{\TT^2}$ with $P_n^{-1}$ as given above, then this condition is not satisfied. However, because of the intrinsic ambiguity on how $\eta$ is defined we may argue that we could always scale $P_n$ by a factor of the form $q^{\frac{(n-1)}{2}}$ where $q$ is a constant (this would be equivalent to scaling $\eta \rightarrow q \eta$). We may then just pick $q$ is such a way as to ensure that $\log P_2=0$. From (\ref{paspn}) we have that $P_2=\sqrt{\frac{\pi}{2}}$. Therefore, by choosing $q=\frac{\pi}{2}$ we may construct a scaled version of $P_n$ given by
\beq 
\tilde{P}_n=\left(\frac{\pi}{2}\right)^{\frac{n-1}{2}}P_n=(\mathcal{C}_{\TT\TT}^{\TT^2})^{-1}
\eeq 
which automatically has the desired properties 
\beq 
\tilde{P}_1=P_1 \quad \text{and} \quad \tilde{P}_2=1.
\eeq 
It is this function $\tilde{P}_n$ which is plotted in Fig.~6 (dashed blue line). The agreement with our data is reasonably good making this identification plausible. It would be nice to have an alternative analytical derivation  of $\mathcal{C}_{\TT\TT}^{\TT^2}$ directly for the non-compactified massless free Boson. For comparison, it is easy to carry out a large $n$ expansion of the function $\log\tilde{P}_n$: the leading terms are $-1.25 + 0.34 n - 0.5 \log n$. The linear term in $n$ is in rather good agreement with the two fits presented in Fig.~6 as they both reproduce well the large $n$ behaviour of our data. The $\log n$ term is  well captured by the second fit. 

\medskip 

In conclusion, our form factor-based numerical and analytical results lead us to conclude that the emergence of a $\log(\log\ell))$ correction in (\ref{log22}) is closely associated with a similar correction in (\ref{log11}). 
Combining our results with the expansion (\ref{xss1}) that follows from \cite{log} and the suggestion coming from \cite{Part2} that the ratio (\ref{tn}) must be finite we propose that the R\'enyi entropy of the non-compatified massless free Boson has the behaviour
\beq 
S_n(\ell):=\frac{\log{\langle \TT(0)\tilde{\TT}(\ell) \rangle_n}}{1-n} =\frac{n+1}{6n}\log\ell+  \log(\log\ell)+O(1), \label{universal}
\eeq 
Since the $\log(\log\ell)$ correction is independent of $n$ the same correction should also contribute to the von Neumann EE.  It is worth noting that the presence of such subleading corrections was missed in the original treatment of the non-compactified massless free Boson \cite{callan}. 

When starting this investigation, the presence of $\log(\log\ell)$ term in the R\'enyi entropy of the non-compactified massless free Boson was entirely unexpected and, as far as we know, had not been suggested by any previous studies. We have now found that a form factor computation combined with various other results strongly suggests its presence. It is worth considering whether or not such a term is universal in the same sense as the leading $\log\ell$ term is. In other words, is the coefficient $+1$ of $\log(\log\ell)$ in (\ref{universal}) a universal number? Based on our understanding to date, there are strong hints that it is not, but that it may depend on the regularisation scheme used. In particular, we understand the the study \cite{Part2} produces a different coefficient, both in sign and absolute value. On the other hand, unpublished numerical studies due to Andrea Coser and Cristiano de Nobili\footnote{We are very thankful to Andrea and Cristiano for finding time in their busy schedule to investigate the result (\ref{universal}) during O.A.C.-A. visit to SISSA in July 2016.} employing an infinitely long harmonic chain and subsystems of sizes varying from few sites to thousands of sites, have found no evidence of such term.
We do not yet understand how these various results can be reconciled but it is something we would like to investigate in the future.

\section{Three Point Couplings and out of Equilibrium Negativity}
Another way of obtaining the value $\lim_{n_e\rightarrow 1^+} \mathcal{C}_{\TT\TT}^{\TT^2}$ is to compare to other existing numerical results. In particular, in \cite{EZ}  the negativity of the harmonic chain, a discrete system whose continuous limit is described by a massless (non-compactified) free Boson, was numerically investigated. In particular, the LN of the harmonic chain out of equilibrium was numerically evaluated. The set up is one in which two harmonic chains are independently thermalized at temperatures $\beta_r^{-1}$ and $\beta_l^{-1}$ and then connected  and let to evolve unitarily (e.g. quenched) at time $t=0$. The time evolution of the LN is then investigated. In this context, the authors obtained very nice numerical results which were later shown to be in full agreement with predictions based on CFT \cite{MB}. Fig.~6 in \cite{EZ} is of particular interest as the ``staircase" pattern of the LN, as well as the height of the steps, have a  CFT interpretation \cite{MB}. In this figure the LN of adjacent regions at the same temperature $1/\beta$ is presented as a function of time $t$ for different choices of $\beta$. Consider equation (90) in \cite{MB}. This equation essentially says the following: if we take the data in Fig.~6 in \cite{EZ} we will see that there is an initial region around $t=0$ where the negativity grows logarithmically as
\beq 
\mathcal{E}_1=\frac{1}{2}\log t + c_1,
\label{re1}
\eeq 
and $c_1$ is a non-universal constant.  Then, the negativity reaches a plateau which is temperature-dependent and is described by formula (88) in \cite{MB}. The value of the negativity at the plateau is given by
\beq 
\mathcal{E}_2=\frac{1}{2}\log \beta + c_2,
\label{re2}
\eeq 
where $c_2$ is another non-universal constant. The key observation made in \cite{MB} is that although (\ref{re1}) and (\ref{re2}) involve non-universal constants, the difference between these constants is a universal CFT quantity related to the three point coupling $\lim_{n_e \rightarrow 1}\mathcal{C}_{\TT \TT}^{\TT^2}$. More precisely
\beq 
-2\log(\lim_{n_e\rightarrow 1^{+}}\mathcal{C}_{\TT\TT}^{\TT^2})=c_1-c_2-\frac{1}{2}\log(2\pi).
\eeq 
From Fig.~6 in \cite{EZ} it is easy to obtain an approximate value of $c_2$ as it is determined by the heights of the first plateau in each curve. Unfortunately we have not had access to the raw data so we could only determine the heights approximately. Considering the four curves in the figure we find that for $\beta=5$ the plateau is located around $\mathcal{E}_2=0.936$, for $\beta=10$ we have a plateau at
$\mathcal{E}_2=1.305$, for $\beta=20$ the plateau is $\mathcal{E}_2=1.645$ and finally, although not very clearly defined the highest point of the curve with $\beta=50$ corresponds to $\mathcal{E}_2=2$. Employing these values we obtain four different predictions for the constant $c_2$. Their average is 
 $c_2=0.119$.

The value of $c_1$ is a bit harder to estimate visually, but it can be obtained by taking a few points on the curves for small $t$. For example, for $\beta=50$ where the logarithmic behaviour is visible for a larger range of values of $t$ we find that for $t=6$ we have $\mathcal{E}_1=1.305$. Similarly, we have $\mathcal{E}_1=1.645$ for $t=11.5$, $\mathcal{E}_1=1.844$ for $t=20$ and $\mathcal{E}_1=0.376$ for $t=15$. Each of these points leads to a value of $c_1$. Taking the average of all four values we find $c_1=0.389$.

With these values we can then propose that 
\beq 
-2\ln(\lim_{n_e \rightarrow 1^+}\mathcal{C}_{\TT\TT}^{\TT^2})=c_1-c_2-\frac{1}{2}\log(2\pi)=-0.645
\eeq
or $\lim_{n_e \rightarrow 1^+}\mathcal{C}_{\TT\TT}^{\TT^2}\approx 1.38$. Given the approximate values we have used and the fact that the results of \cite{EZ} are numerical (not exact), this estimate is in very good agreement with (\ref{exact}). This agreement is also remarkable because the CFT interpretation given in \cite{MB} did not consider the possibility of $\log\log$ corrections to the LN yet, like the form factor approach, it seems to still capture universal information about the CFT. We speculate that the subtraction of the constants $c_1$ and $c_2$ has a similar divergence-canceling effect as described in section 6.

\section{Conclusions and Outlook}
In this paper we have studied the short-distance behaviour of the normalised two-point functions of branch point twist fields $\frac{\bra \TT(0)\tilde{\TT}(\ell) \ket_n}{\bra \TT\ket_n^2}$ and $\frac{\bra \TT(0) {\TT}(\ell) \ket_{n}}{\bra \TT\ket^2_{n}}$ in the replica massive 1+1 dimensional free Boson theory. Our work is based on the use of the form factor approach which allows us to obtain the exact matrix elements of the fields $\TT$ and $\tilde{\TT}$ up to their expectation values. For this reason, it is natural to consider the ratios above, where the dependence on the expectation values is effectively canceled out. From our numerical and analytical results we conclude that
\beqa 
\frac{\bra \TT(0)\tilde{\TT}(\ell) \ket_n}{\bra \TT\ket_n^2} &\sim& \frac{a(n)}{\ell^{b(n)}(\log\ell)^{n-1}}\qquad \text{and} \nonumber\\ 
\frac{\bra \TT(0) {\TT}(\ell) \ket_{n_e}}{\bra \TT\ket^2_{n_e}} &\sim& \frac{c_e(n_e)}{\ell^{d_e(n_e)}(\log\ell)^{\frac{n_e}{2}}}, \qquad \frac{\bra \TT(0) {\TT}(\ell) \ket_{n_o}}{\bra \TT \ket_{n_o}^2} \sim\frac{c_o(n_o)}{\ell^{d_o(n_o)}}, \label{90}
\eeqa  
for $\ell \ll 1$.  The coefficients $b(n), d_{e,o}(n)$ are related to the conformal dimensions of twist fields and known from CFT. On the other hand, the coefficients $a(n)$ and $c_{e,o}(n)$ are universal quantities (ratios) given by
\beq 
a(n)=\bra \TT \ket_n^{-2},\qquad c_{e}(n_e)=\frac{\bra \TT\ket_{\frac{n_e}{2}}^2 \mathcal{C}_{\TT\TT}^{\TT^2}}{\bra \TT \ket_{n_e}^2}\quad \text{and} \quad c_{o}(n_o)=\frac{ \mathcal{C}_{\TT\TT}^{\TT^2}}{\bra \TT \ket_{n_o}}. \label{diver}
\eeq 

In this paper we have shown that a form factor approach can provide extremely accurate predictions for the powers $b(n)$ and $d_{e,o}(n)$. Whereas $b(n)$ may be expressed as an infinite sum of simple terms (\ref{ul}), $d_{e,o}(n)$ may be computed from and exact resummation of a form factor expansion (\ref{xx}). In addition, we have performed the analytic continuation from $n$ integer to $n\geq 1$ and real, so that $b(n)$ and $d_{e,o}(n)$ may be obtained from form factor expansions also for non-integer values of $n$. This provides a powerful test of our approach to the analytic continuation of correlators of twist fields, a problem which is of key importance in their applications to measures of entanglement. 

Remarkably, the form factor approach also allows us to obtain infinite-sum representations for some of the ratios (\ref{diver}). 
Interestingly the sums associated to $a(n)$ and $c_e(n_e)$ turn out to be  divergent, whereas the sum representation of $c_o(n_o)$ is not only rapidly convergent but may also be analytically continued to real $n \geq 1$. In this paper we argue that the divergences we have found may be explained by the presence of the $\log\ell$ powers in the denominators of (\ref{90}).  The presence of similar corrections was noted in studies of the LN of the massless free Boson \cite{negativity2}. For the von Neumann and R\'enyi entropies of one interval, they were implicit in the results of \cite{log} and they have now been independently derived in \cite{Part2}. However, they were missed in the original treatment of the non-compactified massless free Boson \cite{callan}. The presence of $\log\ell$ terms in (\ref{90}) directly leads to the $\log(\log\ell)$ term in (\ref{universal}), that is, it leads to the prediction that all R\'enyi entropies as well as the von Neumann entropy of the non-compactified free Boson should be corrected by a $\log(\log\ell)$ term. This result is surprising and it is worth considering whether or not such extra terms are universal. The evidence we have so far suggest that they are not. It appears that the computations to appear in \cite{Part2} also find a $\log(\log\ell)$ term at criticality, albeit with a different coefficient. At the same time, unpublished numerical studies due to Andrea Coser and Cristiano de Nobili have found no evidence of such terms in an infinite harmonic chain for a wide range of sub-system sizes. We must therefore conclude that $\log(\log\ell)$ corrections to the entanglement entropy are probably non-universal and we still need to understand how and why this is the case. Interestingly, the occurrence of $\log(\log\ell)$ corrections to the R\'enyi and von Neumann EE of a single interval in logarithmic CFTs \cite{gurarie} was shown in \cite{BCDLR} and it would be nice to understand better how the non-compactified massless free Boson falls (or not) within this class of theories.

Despite the fact that $a(n)$ and $c_e(n_e)$ are given by divergent sums in our form factor approach, we have numerically observed that certain combinations of these sums are convergent. In particular, it is possible to obtain convergent expressions for
\beq 
\frac{a(n)}{a(2)^{n-1}}=\frac{\bra \TT\ket_n}{\bra \TT\ket_2^{n-1}},\label{tn2}
\eeq 
and 
\beq 
\frac{a(n_e)c_e(n_e)}{a(n_e/2)}=\mathcal{C}_{\TT\TT}^{\TT^2} \quad \text{for n even}.
\eeq 
These numerical observations (in some cases backed by complementary results \cite{Part2} from other approaches) have allowed us to actually fix the powers of $\log\ell$ in (\ref{90}). 

Employing these convergent series we have obtained numerical values of $\log \mathcal{C}_{\TT\TT}^{\TT^2}$ for $n$ even from $n=2$ to $n=14$. We have then attempted to find a good interpolation of the points obtained that would allow us to find the value of $\lim_{n_e \rightarrow 1^{+}}\mathcal{C}_{\TT\TT}^{\TT^2}$.
We observed that different fits can give very different predictions for this constant (perhaps not-surprisingly as we only had few points). More generally, this provides an instructive example of the difficulty of performing the analytic continuation numerically. Luckily we were able to compare these fits to an analytic prediction from \cite{negativity2}. We have also been able to provide a further numerical estimate of the value $\lim_{n_e \rightarrow 1^+}\mathcal{C}_{\TT\TT}^{\TT^2}$ by combining numerical results for the LN out of equilibrium in a harmonic chain \cite{EZ} and their CFT interpretation \cite{MB}. Remarkably, the value obtained also agrees rather well with the analytic prediction \cite{negativity2}.

In \cite{ourneg} an analytic formula for $\bra \TT \ket_n$ was proposed and therefore it is natural to ask whether or not our results for the ratio (\ref{tn}) are matched by this formula. It turns out that the agreement is poor. There are now indications \cite{Part2} that the formula given in \cite{ourneg} was not correct mainly because the presence of $\log\ell$ corrections to the two-point function $\bra \TT(0) \tilde{\TT}(\ell)\ket_n$ had not acknowledged in the original computation.  We hope that a comparison between our results and those of \cite{Part2} will be possible in the near future. 

The current work has demonstrated that the massive free Boson theory allows for a form factor treatment which we can hardly hope to emulate to interacting theories. This is on account of the simplicity of twist field form factors (and the unusual fact that they are all known). For this reason this is an ideal model for which detailed three- and four-point function form factor computations may be feasible leading to new insights into the properties of the LN and the EE of disconnected regions in gapped systems. These are interesting problems which have not yet been addressed for massive models and we hope to return to them in the future. At the same time, despite it being a non-interacting theory, the massless limit of the massive free Boson is a logarithmic CFT and as a consequence, ``unusual" logarithmic divergences are present in the correlators of branch point twist fields. This gives rise to very interesting structures and in particular to  $\log(\log\ell)$ corrections to the R\'enyi and EE of one interval, in agreement with various predictions \cite{BCDLR, Part2, negativity2, log}. It would be interesting to investigate these unusual corrections further and to establish more rigorously whether or not they are universal and/or numerically observable in some discrete realization of the non-compactified massless free Boson theory.

\medskip 

\paragraph{Acknowledgments:} The authors would like to thank Benjamin Doyon and Olivier Blondeau-Fournier for many enlightening discussions over the past year and for sharing with us their unpublished (and closely related) results \cite{Part2}. O.A. Castro-Alvaredo is grateful to Pasquale Calabrese and Erik Tonni for very useful discussions on their work \cite{negativity2, log} which helped establish its connection to (and agreement with) the present results and for their hospitality during visits to SISSA in April and July 2016. O.A. Castro-Alvaredo would like to thank Andrea Coser and Cristiano de Nobili for all their work on finding numerical evidence for the presence of $\log(\log\ell)$ corrections to the entanglement entropy of the harmonic chain. Their negative result (that is, no evidence of such corrections) provides crucial support for the conclusion that such terms are most likely non-universal. Finally, O.A. Castro-Alvaredo is also grateful to the Isaac Newton Institute (Cambridge) for hospitality during the scientific programme {\it Mathematical Aspects of Quantum Integrable Models in and out of Equilibrium}, January 2016, where some preliminary results were presented and useful discussions took place. 

\appendix 
\section{Summation Formulae for the free Boson Theory}

In \cite{ourneg} several summation formulae involving two-particle form factors were obtained for generic 1+1-dimensional QFTs. Let $f(\theta;n)$ be the two-particle form factor as defined in (\ref{2pff}), then these formulae specialize as follows to the massive free Boson theory
\beqa 
\sum_{j=0}^{n-1} f(-x+2 \pi i j;n) f(y+2\pi i j;n)=-\frac{i}{2}\frac{\sinh\left(\frac{x+y}{2}\right)}{\cosh\frac{x}{2}\cosh\frac{y}{2}}\left(f(x+y+ i\pi;n)-f(x+y-i\pi;n)\right).\label{f3}
\eeqa 
The identity above is obtained by analytic continuation in $n$ using a ``cotangent trick'' to turn the sum into a contour integral and then use the kinematic singularities of the function $f(\theta;n)$ to explicitly compute this integral. 
The formula (\ref{f3}) can be easily generalised (by induction) to include additional sums. A similar procedure was employing in \cite{nexttonext} for the free Fermion theory. The formulae for the free Boson are almost identical, up to a few sign changes. The case of interest here corresponds to performing and odd number of sums. Employing once more the notation $x^{j}:=x+ 2\pi i j$, we find
\beqa 
&&\sum_{p_1,\cdots,p_{2j-1}=0}^{n-1} f((-x_1)^{p_1};n)f(x_2^{p_1-p_2};n)\cdots f(x_{2j-1}^{p_{2j-2}-p_{2j-1}};n)f(x_{2j}^{p_{2j-1}};n)= \frac{2i \sinh(\frac{1}{2}\sum_{i=1}^{2j} x_i)}{\prod_{i=1}^{2j} 2 \cosh \frac{x_i}{2}}\nonumber\\
&& \times \sum_{p=1}^j (-1)^p \left(\begin{array}{c}
2j-1\\
j-p
\end{array}\right)\left[f(\sum_{i=1}^{2j} x_i+(2p-1)i\pi;n)-f(\sum_{i=1}^{2j} x_i-(2p-1)i\pi;n) \right]. \label{key}
\eeqa 
An important property of (\ref{key}) is its behaviour in the limit $\sum_{i=1}^{2j} x_i=0$. Although the presence of the sinh-function in the numerator suggests the sum should be zero, this is not the case as the sum in $j$ develops kinematic poles in the same limit. Those contributions arise in three particular instances of the sum: First, when $p=1$ second, when $p= k n$, and thirdly when $p=kn+1$ with $k \in \mathbb{Z}$. In each case a kinematic pole is captured. How many of these poles contribute to the sum depends therefore on the relative values of $p$ and $n$. In general we may write that:
\beqa 
&&\lim_{\sum_{i=1}^{2j} x_i\rightarrow 0}\sum_{p_1,\cdots,p_{2j-1}=0}^{n-1} f((-x_1)^{p_1};n)f(x_2^{p_1-p_2};n)\cdots f(x_{2j-1}^{p_{2j-2}-p_{2j-1}};n)f(x_{2j}^{p_{2j-1}};n)\nonumber\\
&&\qquad\qquad \qquad=h(j,n)\,\,\text{sech}\left(\frac{\sum_{p=2}^{2j}x_p}{2}\right) \prod_{p=2}^{2j} \text{sech}\frac{x_p}{2}.\label{magic}
\eeqa 
with
\beq 
h(j,n):=\frac{1}{2^{2j-1}}\left(\left(\begin{array}{c}
2j-1\\
j-1
\end{array}\right) + \sum_{p=1}^{[\frac{j}{n}]} (-1)^{p n} 
\left(\begin{array}{c}
2j\\
j-p n
\end{array}\right)  \right),
\eeq 
thus, for $n>j$ only the first term contributes. In many computations it will be important to analytically continue $h(j,n)$ from $n$ even or $n$ odd. It is natural to define
\beq 
h^e(j,n)=\frac{1}{2^{2j-1}}\left(\left(\begin{array}{c}
2j-1\\
j-1
\end{array}\right) + \sum_{p=1}^{[\frac{j}{n}]} 
\left(\begin{array}{c}
2j\\
j-p n
\end{array}\right)  \right), \label{heven}
\eeq 
and
\beq 
h^o(j,n)=\frac{1}{2^{2j-1}}\left(\left(\begin{array}{c}
2j-1\\
j-1
\end{array}\right) + \sum_{p=1}^{[\frac{j}{n}]}  (-1)^p
\left(\begin{array}{c}
2j\\
j-p n
\end{array}\right)  \right), \label{hodd}
\eeq 
to be the analytic continuations of $h(j,n)$ from $n$ even and from $n$ odd, to $n$ real and positive, respectively. 

\section{Numerical Precision}
Many of the results of this project are, at least in part, based on the use of numerical algorithms and different types of approximations. For this reason we think it is important to give a brief discussion of how we think these procedures affect the precision of our results. From the scattering theory point of view, we have dealt with the simplest theory we could possibly imagine: non-interacting particles with two-particle scattering matrix $S(\theta)=1$.  In that respect, the number and nature of the challenges we have faced when attempting to evaluate physical quantities numerically has been somewhat surprising. There were two main challenges:
\begin{itemize}
\item[1)] {\it Physical quantities are given in terms of (infinite) slowly convergent sums}: this is particularly striking for the series (\ref{ul}) and for its analytically continued version (\ref{anul}). However, it also emerges less visibly in the series representations (\ref{sum}). In this case, the slow convergence is less apparent because we are able to perform all sums analytically, however it is easy to see that convergence of this sequence is also slow. In all cases, it can be shown that the terms in the series involved decay roughly as $1/j^s$ for $s>1$.
\item[2)] {\it Physical quantities are given in terms of divergent sums}: in some cases, the situation is even worse than 1) and we actually have divergent representations for quantities of physical interest such as the expectation value (\ref{kttt}) and the ratio (\ref{ke}). In fact, we are not aware of any other form factor calculation where such sequences emerge. As discussed in the paper, we have discovered that convergent sequences may still be constructed by combining several divergent ones in a physically meaningful way.  
\end{itemize}
Interestingly, although the free Fermion theory which was studied in \cite{nexttonext} is in many ways very similar in structure and in the sort of computations involved, it turns out that none of these issues arise. We believe that divergence and poor convergence of for factor expansions are both related to the presence of $\log\ell$ divergences in the correlators of the massless free Boson, a feature which is of course absent for the free Fermion (and most other theories of interest). The cancellation of divergences by combining several divergent sum is then equivalent to the statement that although some correlators diverge as $r^{-x} (\log\ell)^y$ for some $x,y>0$ the $\log\ell$ divergences can be canceled by computing instead ratios of several correlators. 

For those sequences which are convergent and which we could evaluate, we used a range a methods that allowed us to sum if not the full sequence at least a large number of terms. The only sequence which we could sum exactly was (\ref{sum}).  A particularly neat example is the sequence (\ref{ktt}) for $n$ odd. For all the values of $n$ we considered, the sum has fully converged after 200 terms (which is fast, compared for example to (\ref{ul}) and (\ref{anul})). However, even to evaluate this sequence we have employed the fit (\ref{fit}), meaning that the values we summed are the result of approximating the integrals (\ref{vej}) by a fit (rather than evaluating (\ref{vej}) for all $j$ up to 200). Instead the integrals (\ref{vej}) were evaluated using a Monte Carlo algorithm for $j=1,\ldots,12$ (these are the dots in Fig.~4) and then a numerical fit of $v_j$ was performed based on these first 12 values. The error on the values depicted in Fig.~5 which is induced by this procedure is hard to estimate, although based on the errors derived from the Monte Carlo (which are very small) and the errors of the fitting $v^{\text{fit}}_j$, which are also small, we expect the results in Fig.~5 to be very accurate. 

This same technique of using fits to be able to sum further terms has been employed in our evaluation of the structure constant $\mathcal{C}_{\TT\TT}^{\TT^2}$ for $n$ even in (\ref{cancel}). Our objective was to find an appropriate fit of those values that would allow us to carry out the numerical extrapolation to $n=1$. As shown in Fig.~6 we were only partly successful in this. Indeed, various sensible interpolating functions turned out to exhibit very different behaviours precisely around $n=1$. This is due to a large extent to the fact that, unfortunately we only had very few values to fit. This in turn is due to the difficulty of finding accurate values for the integrals $u_j(n)$ through Monte Carlo for large $j$. The numerical values depicted in Fig.~6 were obtained by summing (\ref{cancel}) and employing interpolating functions for the integrals $v_j$ and $u_j(n)$. Again, it is hard to estimate the percentage of error induced by employing these fits, mainly because the functions (\ref{fituj}) where obtained by interpolating with only few points (see Fig.~3). 

There is another approximation involved in evaluating (\ref{cancel}) and that is the fact that the divergent parts of the three sums involved do not (numerically speaking) cancel exactly. For all values of $n$ we find that the divergent terms in the sum, which diverge as $1/j$, have coefficients of the order of $10^{-2}$ (which are neglected to achieve convergence). 

Let $\kappa_n$ be the percentage error associated with the truncation and cancellation of divergences in the sum (\ref{cancel}) for each value of $n$. In order to extrapolate to $n=1$ we employed a fit of the form:
\begin{eqnarray}\label{eq:fit}
-\log \mathcal{C}_{\TT\TT}^{\TT^2} \;\; \equiv \;\; y^{\text{fit}}(n) &=& a+bn+\frac{c}{n}
\end{eqnarray}
Let $y(n)$ be the numerical values to be fitted. We define the error $\sigma(n)=\kappa_n y(n)$ where $\kappa_n$ is the percentage error on the value $y(n)$ stemming from the various approximations discussed above (e.g if the error on the $y(n)$ were of 10\% then $\kappa_n=0.1$). In general we would expect that error to be approximately the same for every $n$. However, in the case of the data in Fig.~6 we have seen that the value $y(2)=0$ is actually exact (see eq.~(\ref{goodsum})). We may assume that all $\kappa_n$ are the same for $n>2$ and $\kappa_2=0$. In order to compute the values (and their relative errors) of the fitting constants $a$, $b$ and $c$ we perform a least-squared fitting \cite{TaylorBook}. In particular, the constants $a$, $b$, and $c$ are such that minimise the quantity:
\begin{eqnarray}\label{eq:chi}
\chi^2 &=& \sum_{n \,\, \text{even}} \frac{(y^{\text{fit}}(n)-y(n))^2}{\sigma(n)^2},
\end{eqnarray}
 Taking derivatives with respect to the coefficients $a$, $b$, and $c$ we can minimise $\chi^2$:
\begin{equation}
\left\{
\begin{array}{ccccc}
\frac{1}{2}\frac{\partial \chi^2}{\partial a} &=& \sum\limits_{n \,\, \text{even}} \frac{y^{\text{fit}}(n)-y(n)}{\sigma(n)^2} &=& 0 \\
\frac{1}{2}\frac{\partial \chi^2}{\partial b} &=&  \sum\limits_{n \,\, \text{even}} \frac{n(y^{\text{fit}}(n)-y(n))}{\sigma(n)^2} &=& 0\\
\frac{1}{2}\frac{\partial \chi^2}{\partial c} &=& \sum\limits_{n \,\, \text{even}} \frac{y^{\text{fit}}(n)-y(n)}{n\sigma(n)^2}&=& 0
\end{array}
\right.
\end{equation}
The above system can be easily solved for $a$, $b$, and $c$:
\begin{equation}\label{eq:abc}
\begin{array}{ccl}
a &=& \frac{1}{\Delta}\sum\limits_{n \,\, \text{even}}\sum\limits_{m \,\, \text{even}}\sum\limits_{p \,\, \text{even}}\frac{p}{\sigma(n)^2\sigma(m)^2\sigma(p)^2}\left(1 -\frac{m^2}{n^2}-\frac{p}{n}\right)\left(\frac{y(p)}{p}-\frac{y(m)}{m}\right),\\
b &=& \frac{1}{\Delta}\sum\limits_{n \,\, \text{even}}\sum\limits_{m \,\, \text{even}}\sum\limits_{p \,\, \text{even}}\frac{y(p)}{\sigma(n)^2\sigma(m)^2\sigma(p)^2}\left[\frac{m}{n^2}\left(1-\frac{n^2}{m^2}\right)+\left(\frac{p}{m n}-\frac{1}{p}\right)\left(\frac{n}{m}-1\right)\right],\\
c &=& -\frac{1}{\Delta}\sum\limits_{n \,\, \text{even}}\sum\limits_{m \,\, \text{even}}\sum\limits_{p \,\, \text{even}}\frac{y(p)}{\sigma(n)^2\sigma(m)^2\sigma(p)^2}\left(1-\frac{m}{n}\right)\left(m-p-\frac{m n}{p}\right),\\
\Delta &=& \sum\limits_{n \,\, \text{even}}\sum\limits_{m \,\, \text{even}}\sum\limits_{p \,\, \text{even}}\frac{1}{\sigma(n)^2\sigma(m)^2\sigma(p)^2}\left(\frac{mp}{n^2}-\frac{2p}{n}-\frac{p^2}{n^2}+\frac{1}{nm}+1\right).
\end{array}
\end{equation}
Since only the $y(n)$ values are affected by error, the associated error of the $a$, $b$, and $c$ constants is given by:
\begin{equation}\label{eq:abc:err}
\left\{
\begin{array}{ccc}
\delta a &=& \sqrt{\sum\limits_{n \,\, \text{even}}\left(\frac{\partial a}{\partial y(n)}\sigma(n)\right)^2}\\
\delta b &=& \sqrt{\sum\limits_{n \,\, \text{even}}\left(\frac{\partial b}{\partial y(n)}\sigma(n)\right)^2}\\
\delta c &=& \sqrt{\sum\limits_{n \,\, \text{even}}\left(\frac{\partial c}{\partial y(n)}\sigma(n)\right)^2}
\end{array}
\right.
\end{equation}
Using equations (\ref{eq:abc}) and (\ref{eq:abc:err}) we can compute the numerical values of the fitting coefficients:
\begin{eqnarray}
a = -0.990987... \quad 
b = 0.266514... \quad \text{and} \quad
c = 0.915919...
\end{eqnarray}
and their errors. Assuming for instance that $\kappa_n=0.1$ for all $n>2$ (that is a 10\% error on each of the numerical values $y(n)$) we obtain errors:
\beq 
\delta a=0.157838...\quad 
\delta b = 0.0201026... \quad \text{and} \quad
\delta c = 0.240441...
\eeq 
And the value of $\mathcal{C}_{\TT\TT}^{\TT^2}$ in $n=1$ is then given by:
\begin{eqnarray}
\lim_{n_e\rightarrow 1^+} \mathcal{C}_{\TT\TT}^{\TT^2}&=& e^{-(a+b+c)} \;\;=\;\; 0.77 \pm 0.51 \label{result}
\end{eqnarray}
where the error at $n_e=1$ is given by:
\begin{eqnarray}
\delta \mathcal{C}_{\TT\TT}^{\TT^2} &=& e^{-(a+b+c)}(|\delta a|+|\delta b |+|\delta c|)= 0.506659...
\end{eqnarray}
A similar study could be performed for the other fit presented in Fig.~6. For a fit of the form $y^{\text{fit}}(n)=a + b n + c\log n$ and assuming again $\kappa_n=0.1$ for $n>2$ we obtain
\beq 
a=-0.308\pm 0.028, \quad 
 b =0.312 \pm 0.031, \quad \text{and}\quad 
 c=-0.456\pm 0.119.
\eeq 
giving
\beq
\lim_{n_e\rightarrow 1^+} \mathcal{C}_{\TT\TT}^{\TT^2}= 1.00\pm 0.06.
\eeq 
As we can see not only the second fit was in better agreement with the analytical prediction around $n=1$ but it entails in general a much smaller error. 
\bibliographystyle{phreport}

\end{document}